\documentclass[twocolumn,astrosymb,trackchanges]{aastex701}

\usepackage{xspace}

\newcommand\kms{\ensuremath{{\rm km}~{\rm s}^{-1}}\xspace}
\newcommand\angstrom{\ensuremath{\mathring{\rm A}}\xspace}
\newcommand\ergs{\ensuremath{{\rm erg}~{\rm s}^{-1}}\xspace}
\newcommand\ergscm{\ensuremath{{\rm erg}~{\rm s}^{-1}~{\rm cm}^{-2}}\xspace}
\newcommand\msun{\ensuremath{M_\odot}\xspace}
\newcommand\msunyr{\ensuremath{M_\odot~{\rm yr}^{-1}}\xspace}

\newcommand\kmsMpc{\ensuremath{{\rm km}~{\rm s}^{-1}~{\rm Mpc}^{-1}}\xspace}

\newcommand\ovi{\ion{O}{6}\xspace}
\newcommand\oviwave{\ion{O}{6}~$\lambda\lambda$1032,1038\,\angstrom\xspace}

\newcommand\sfr{\ensuremath{{\rm SFR}}\xspace}
\newcommand\mdotovi{\ensuremath{\dot{M}_{\rm O\,VI}}\xspace}
\newcommand\mdotx{\ensuremath{\dot{M}_{\rm X,spec}}\xspace}
\newcommand\mdothcf{\ensuremath{\dot{M}_{\rm X,spec,hcf}}\xspace}

\newcommand\mdotmax{\ensuremath{\dot{M}_{\rm max}}\xspace}
\newcommand\mdotmaxap{\ensuremath{\dot{M}_{\rm max,ap}}\xspace}

\usepackage{threeparttable}
\usepackage{amsmath}
\usepackage{amssymb}
\usepackage{amsfonts}
\usepackage{mathbbol}
\usepackage{physics}
\usepackage{gensymb}
\usepackage{tabularx}
\usepackage{fontawesome}
\usepackage{hyperref}
\usepackage{xurl}

\let\tablenum\relax
\usepackage{siunitx}

\shorttitle{\ovi Emission in Starbursting BCGs}
\shortauthors{Reefe et al.}

\graphicspath{{./}{figures/}}

\begin{document}

\title{\ovi Emission in Starbursting BCGs: An HST-COS Extension of ``Reigniting the FUSE''}

\correspondingauthor{Michael Reefe}
\email{mreefe@mit.edu}

\author[0000-0003-4701-8497]{Michael Reefe}
\altaffiliation{National Science Foundation, Graduate Research Fellow}
\affiliation{Kavli Institute for Astrophysics \& Space Research, Massachusetts Institute of Technology, Cambridge, MA 02139, USA}
\email{mreefe@mit.edu}

\author[0000-0001-5226-8349]{Michael McDonald}
\affiliation{Kavli Institute for Astrophysics \& Space Research, Massachusetts Institute of Technology, Cambridge, MA 02139, USA}
\email{mcdonald@space.mit.edu}

\author[0000-0002-8823-0606]{Marios Chatzikos}
\affiliation{Department of Physics \& Astronomy, University of Kentucky, Lexington, KY 40506, USA}
\affiliation{Eureka Scientific, 2452 Delmer Street Suite 100, Oakland, 94602, CA, USA}
\email{mchatzikos@gmail.com}

\author[0000-0002-2808-0853]{Megan Donahue}
\affiliation{Department of Physics and Astronomy, Michigan State University, East Lansing, MI 48824, USA}
\email{donahu42@msu.edu}

\author[0000-0002-9378-4072]{Andrew C. Fabian}
\affiliation{Institute of Astronomy, University of Cambridge, Madingley Road, Cambridge, CB3 0HA, UK}
\email{acf@ast.cam.ac.uk}

\author[0000-0003-2754-9258]{Massimo Gaspari}
\affiliation{Department of Physics, Informatics and Mathematics, University of Modena and Reggio Emilia, 41125 Modena, Italy}
\email{massimo.gaspari@unimore.it}

\author[0000-0002-3031-2326]{Eric Miller}
\affiliation{Kavli Institute for Astrophysics \& Space Research, Massachusetts Institute of Technology, Cambridge, MA 02139, USA}
\email{milleric@space.mit.edu}

\author[0000-0001-6638-4324]{Valeria Olivares}
\affiliation{Departamento de F\'isica, Universidad de Santiago de Chile, Av. Victor Jara 3659, Santiago 9170124, Chile}
\affiliation{Center for Interdisciplinary Research in Astrophysics and Space Exploration (CIRAS), Universidad de Santiago de Chile, Santiago
9170124, Chile}
\email{valeria.olivares@usach.cl}

\author[0000-0003-3521-3631]{Taweewat Somboonpanyakul}
\affiliation{Department of Physics, Faculty of Science, Chulalongkorn University, 254 Phayathai Road, Pathumwan, Bangkok 10330, Thailand}
\email{taweewat.s@chula.ac.th}

\author[0000-0002-3514-0383]{G. Mark Voit}
\affiliation{Department of Physics and Astronomy, Michigan State University, East Lansing, MI 48824, USA}
\email{voit@msu.edu}


\begin{abstract}

In this extension of our ``Reigniting the \textit{FUSE}'' series, we have obtained new far-ultraviolet spectroscopy of 3 starbursting brightest cluster galaxies with the Cosmic Origins Spectrograph (COS) aboard \textit{HST}.  With these new data, in combination with archival COS data of 2 additional systems, we measure star formation rates (SFRs) from the young stellar continuum and cooling rates through $10^{5.5}$ K gas with the \oviwave doublet (\mdotovi).  We use these measurements to increase the size of the ``Reigniting the \textit{FUSE}'' sample, which itself contains high resolution far UV spectroscopy for 29 nearby massive elliptical galaxies from \textit{FUSE}, and extend it 1 dex upwards in SFR and cooling rate.  Our derived SFRs agree well with photometric \textit{GALEX}$+$\textit{WISE} SFRs from the literature. The new data also reinforce the previously observed trend in cooling efficiency (\mdotovi/\mdotmax; where \mdotmax is the X-ray luminosity-based cooling rate) versus mass, with groups cooling more efficiently than clusters by a factor of $\sim$4--5.  When we isolate a sample of the most star-forming central galaxies, we find that the star formation rates, X-ray spectroscopic cooling rates, and UV spectroscopic cooling rates are all consistent with one another, on average, while retaining substantial system-to-system scatter, reflecting time-depenent multiphase condensation and cold gas reservoir evolution.  The remainder of the sample, in contrast, overwhelmingly shows ratios of $\sfr/\mdotovi \sim \sfr/\mdotx \sim 10\%$.

\end{abstract}

\section{Introduction} \label{sec:intro}

Brightest Cluster Galaxies (BCGs) sit at the centers of the most massive halos in the universe, acting as a gravitational focus for the hot ($>10^7$\,K) and diffuse intracluster medium (ICM), which fills the cluster volume and outweighs the stars by an order of magnitude. In $\sim 1/3^{\rm rd}$ of these systems (so-called ``cool cores''), this gas is dense enough that it should cool and condense onto the BCG, forming new stars at rates of 100s--1000s of \msunyr \citep{1994ARA&A..32..277F}, which is inferred by the luminosity and temperature of the cooling gas \citep[$\mdotmax \propto L_X/kT_X$;][]{1988xrec.book.....S}.  More often than not, this cooling is balanced by feedback from an active galactic nucleus, or AGN \citep[see reviews by][]{2007ARA&A..45..117M, 2012ARA&A..50..455F, 2012NJPh...14e5023M, 2020NatAs...4...10G}, lowering the actual star formation rates (SFRs) to $\lesssim 10$ \msunyr \citep{1985ApJS...59..447H, 1987MNRAS.224...75J, 1989ApJ...338...48H, 1989AJ.....98.2018M, 1999MNRAS.306..857C, 2008ApJ...681.1035O, 2010ApJ...721.1262M, 2012ApJS..199...23H, 2018ApJ...858...45M, 2023ApJ...947...44C}.  In this picture, \mdotmax represents the maximum average cooling rate over long periods, as it does not account for feedback and heating effects.  Therefore, directly probing the cooling rate of the gas at cooler temperatures may reveal rates much lower than \mdotmax.  In an idealized, steady-state cooling flow, these cooling rates should agree at all temperatures and be roughly comparable to the \sfr.

Such measurements can be made by converting the luminosities of high-ionization emission lines, sensitive to particular temperatures, into cooling rates (for example with soft X-ray spectroscopy of \ion{O}{7} and \ion{O}{8}, \mdotx, or the far-UV line \oviwave, \mdotovi).  \mdotx and \mdotovi are inherently model-dependent quantities, assuming a single-pass steady cooling flow, and are not actual measurements of a net mass deposition rate. Nevertheless, they are useful for comparisons with \mdotmax and the \sfr and distinguishing different modes of cooling and feedback.  Interestingly, \mdotovi, \mdotx, and \sfr most often do not agree with each other, even when accounting for statistical and systematic uncertainties \citep{2016A&A...595A.123M}.  Soft X-ray lines recover \mdotx far below \mdotmax, and occasionally below the \sfr \citep{2001A&A...365L..99K, 2001A&A...365L..87T, 2001ApJ...560..194M, 2001A&A...365L.104P, 2003ApJ...590..207P, 2006PhR...427....1P}, while \mdotovi typically exceeds both \mdotx and the \sfr by an order of magnitude  \citep{2001ApJ...560..187O, 2004A&A...421..503L, 2005ApJ...635.1031B, 2006ApJ...642..746B, 2014ApJ...791L..30M, 2015ApJ...811..111M, 2017ApJ...835..216D, 2026arXiv260913368R}.  To explain this, three explanations have been proposed: \ovi is enhanced by non-radiative processes such as thermal conduction or turbulent mixing \citep{1993ApJ...405L..17C, 1990MNRAS.244P..26B, 1993ApJ...407...83S}; soft X-ray lines are suppressed by self-absorption in a ``hidden'' cooling flow \citep{2022MNRAS.515.3336F, 2023MNRAS.521.1794F, 2023MNRAS.524..716F, 2024MNRAS.535.2173F}; and/or the SFRs themselves are underestimated, perhaps due to non-standard initial mass functions (IMFs).

Modern hydrodynamic simulations reveal the highly stochastic behavior of cooling over time \citep{2013MNRAS.432.3401G}.  Condensation from the hot atmosphere, cloud uplift and recycling, molecular reservoir build-up, star formation, and inward transport are all distinct stages of the baryon cycle.  Jet-regulated chaotic cold accretion (CCA) simulations show that the production of cold gas can become decoupled from its subsequent transport to the supermassive black hole \citep{2026arXiv260527511C, 2026arXiv260527508P}.  Therefore, various measures of cooling and star formation need not agree instantaneously in any individual system.  When looking at the population as a whole, this effect may increase system-to-system scatter and/or introduce systematic offsets if tracers have sensitivities to different times/phases in the baryon cycle.  This also links to the larger-scale coupling between halo regulation and nuclear feeding \citep{2017ApJ...837..149G}.

This paper is an extension of our ``Reigniting the FUSE'' series (Paper I: \citet{2026arXiv260905675R}; Paper II: \citet{2026arXiv260913368R}), in which we have performed a homoegenous re-analysis of archival \textit{Far Ultraviolet Spectroscopic Explorer} (\textit{FUSE}) data of 29 massive elliptical galaxies.  In Paper I, we analyzed the stellar populations, measured SFRs, and found based on the stellar metallicities that the youngest stars cannot have formed out of recycled gas from the old stellar populations.  In Paper II, we measured cooling rates from the \ovi, along with \mdotmax and \mdotx.  The relationship between \mdotovi and \mdotmax showed that cooling may be more efficient in groups, and less so in clusters, owing to differences in the physical scale of AGN feedback between the group and cluster scales.  Meanwhile, \mdotovi and \sfr are highly correlated but offset from each other when assuming pure cooling models.  Hidden cooling flow models can bring \mdotx and \mdotovi into agreement (at the expense of \sfr), while turbulent mixing layers models (with fiducial parameters on gas viscosity and conductivity) and bottom-heavy IMFs can bring \mdotovi into agreement with the \sfr (at the expense of \mdotx).

The principal limitation of these studies is the dynamic range.  The \textit{FUSE} targets are mostly nearby, quiescent ellipticals, with SFRs of at most a few \msunyr and correspondingly modest \mdotmax.  The most active galaxy in this sample, which itself is an outlier, is NGC 1275, the BCG of the Perseus Cluster, with an \sfr of $34^{+10}_{-8}$ \msunyr.  This excludes the extreme cool core systems at the top of the mass function, harboring starbursting BCGs that are arguably the best laboratories for studying cooling (e.g. Abell 1835: \citet{2006ApJ...648..164M}, the Phoenix Cluster: \citet{2012Natur.488..349M}).  In a vacuum, these systems would make valuable additions to this survey solely based on the increase in dynamic range.  However, there is the additional benefit of testing whether their apparent increase in efficiency (\sfr/\mdotmax) is attributable to an increase in \mdotovi/\mdotmax or an increase in \sfr/\mdotovi.

In this paper we therefore incorporate far-UV spectroscopy of 5 starbursting BCGs obtained with the Cosmic Origins Spectrograph (COS) aboard the \textit{Hubble Space Telescope} (\textit{HST}), of which 3 are newly acquired specifically for this study.  We adopt the methodology of Papers I and II essentially unchanged, and refer the reader there for details.  $\S$\ref{sec:sample} describes the sample and observations, $\S$\ref{sec:methods} summarizes the analysis and notes the few departures from Papers I and II, $\S$\ref{sec:results} presents the SFRs, \ovi\ fluxes, and cooling rates alongside the \textit{FUSE} sample, and $\S$\ref{sec:discuss} discusses what the high-mass end implies for the cooling flow problem.  Throughout, we assume a flat $\Lambda$CDM cosmology with $H_0=70$ \kmsMpc, $\Omega_m = 0.27$, and $\Omega_\Lambda = 0.73$.  Uncertainties are $1\sigma$ (68\% confidence) unless otherwise stated.

\section{Sample} \label{sec:sample}

\begin{deluxetable*}{lDDDcDL}
\def\arraystretch{0.85}
\tabletypesize{\footnotesize}
\tablecaption{Target observational information for our sample.}
\label{tab:info}
\tablehead{
    \colhead{Target Name} & \twocolhead{R.A. (deg.)} & \twocolhead{Dec. (deg.)} & \twocolhead{Redshift} & \colhead{Observing Setup} & \twocolhead{$t_{\rm exp}$ (s)} & \colhead{Pointings} 
}
\decimals
\startdata
\multicolumn{11}{c}{New Observations}\\
\hline
Abell 1835           & 210.25842   & 2.87862    & 0.251958  & G130M/1222 & 12132 & 1 \\
CHIPS 1911+4455      & 287.75417   & 44.92222   & 0.485     & G160M/1600 & 31209 & 1 \\
MACS J1931.8-2634    & 292.95000   & -26.58333  & 0.352     & G130M/1291 & 12370 & 1 \\
\hline
\multicolumn{11}{c}{Archival Observations}\\
\hline
RX J1532.9+3021      & 233.22407   & 30.34976   & 0.362049  & G140L/1280 & 12076 & 1 \\
Phoenix              & 356.18295   & -42.72012  & 0.597320  & G160M/1577 & 19751 & 2 \\
\hline
\enddata
\end{deluxetable*}

This \textit{HST}-COS sample extends our 29-galaxy \textit{FUSE} sample to significantly higher redshift, higher mass, and higher star formation rates. Targets were selected based on 3 criteria: (1) no evidence for an optically bright AGN; (2) published SFRs $> 100\,\msunyr$; and (3) redshift high enough that the COS aperture subtends $> 10$ kpc in diameter, but low enough that COS retains sufficient sensitivity ($0.25 \lesssim z \lesssim 0.7$).  This resulted in 3 BCGs being chosen: Abell 1835, CHIPS 1911+4455 (hereafter CHIPS 1911), and MACS J1931.8-2634 (hereafter MACS 1931).  These data are collected as a part of program ID 17716 (PI: McDonald).  

The observing strategy uses the COS/FUV spectrograph with medium resolution gratings and central wavelength setups that vary with each source according to its redshift to allow for the observation of the \oviwave doublet alongside the young stellar continuum.  Each target is observed at a sufficient depth to detect \ovi at $3\sigma$ confidence, assuming conservatively that $\mdotovi = \sfr$.  To point precisely at each target, COS's standard target acquisition routines, which are optimized for point sources, are performed on nearby UV-bright stars, after which the telescope is slewed by a fixed coordinate offset to position the extended \ovi emission within COS's $2.5''$ diameter circular aperture.

\begin{figure*}
    \centering
    \includegraphics[width=\linewidth]{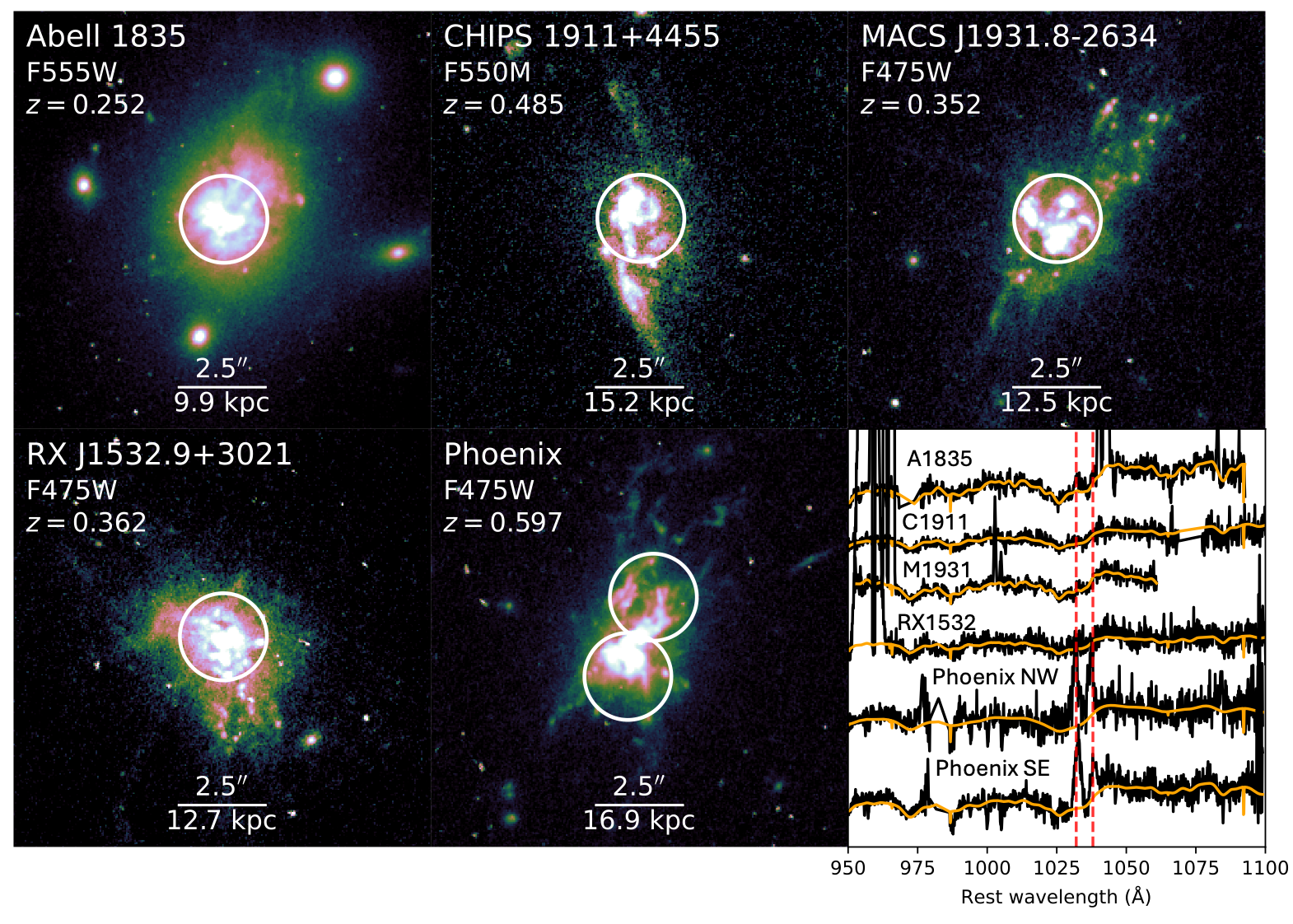}
    \caption{Images of each system with \textit{HST}-COS data are shown, using data from \textit{HST}-ACS. Rest-frame blue-optical/near-UV filters are chosen, based on the availability of data for each system.  The position of the $2.5''$ diameter COS aperture is shown, which contains the majority of the clumpy star formation in all cases.  In the bottom-right panel, the COS spectra of each system are shown in the rest frame (in order from top to bottom: Abell 1835, CHIPS 1911+4455, MACS J1931.8-2634, RX J1532.9+3021, Phoenix NW, Phoenix SE), along with their best-fitting stellar continuum models in orange.  The fluxes have been arbitrarily scaled and shifted for visual clarity.  The positions of the \ovi doublet are marked by the red dashed vertical lines.}
    \label{fig:cos_apertures}
\end{figure*}

Alongside these new observations, we also include 2 archival \textit{HST}-COS observations of RX J1532.9+3021 (hereafter RX 1532) and the Phoenix Cluster. These are obtained from program ID 13367 \citep[PI: Donahue;][]{2017ApJ...835..216D} and 13456 \citep[PI: McDonald;][]{2015ApJ...811..111M}, respectively.  Both of these targets fulfill all of our selection criteria above, with the exception of Phoenix hosting a bright quasar, which we treat with special attention in our analysis. A summary of the observational setups and exposure times of all new and archival data is presented in Table \ref{tab:info}.  The position of the COS aperture is shown over rest-frame blue-optical/near-UV broadband images from \textit{HST}'s Advanced Camera for Surveys (ACS) in Figure \ref{fig:cos_apertures}, alongside the COS FUV spectra.  For details on the 29 \textit{FUSE} systems, we direct the reader to Paper I.

\section{Methods} \label{sec:methods}

\subsection{Data Reduction}

For each system, all COS/FUV exposures are coadded into a single deep spectrum, weighting by the inverse variance in each pixel.  The spectra are then binned to a wavelength sampling of $0.2~\angstrom$, matching the binned resolution used in Papers I and II.  This improves the $S/N$ per pixel, while maintaining a fine enough sampling to resolve the typical scale of gas motions in the ICM ($\sim 60~\kms$).  Backgrounds, which are present due to a combination of detector dark currents, earthshine, zodiacal light, and airglow, are saved separately from the data so that they may be included as a free model component in our fitting procedures.

Following Paper I, we aim to obtain aperture-matched broadband photometry at longer wavelengths to anchor our stellar population models.  The sky surveys utilized in Paper I (GALEX/SDSS/DES/2MASS/WISE), while ideal for the nearby, bright targets with \textit{FUSE} data, are less than ideal for the COS targets in this study.  For one, these surveys are shallow, and these targets are at much higher redshifts ($z \gtrsim 0.2$) and suffer from cosmic surface brightness dimming $\propto (1+z)^4$.  More importantly, however, is the spatial resolution:  COS's aperture size of $2.5''$ is much smaller than \textit{FUSE}'s LWRS aperture of $30'' \times 30''$, and is comparable to or smaller than the PSF size in the longer wavelength surveys.  This would require a large and uncertain aperture correction factor to be applied to any photometry obtained.

Therefore, we instead opt to use photometry from \textit{HST} imaging, using the Advanced Camera for Surveys (ACS) and the Wide Field Camera 3 (WFC3).  Archival imaging data exists for all 5 of our targets in at least 2 bands, with some having coverage in up to 18 bands.  We extract photometry from each available band within the $2.5''$ COS aperture and synthesize UV-to-IR SEDs for each system.  Statistical errors on the photometry are propagated from the inverse variance images provided by the \textit{HST} imaging, while for systematic errors we assume a conservative 2\% absolute flux calibration uncertainty.

\subsection{Calculating Cooling Rates}

After assembling the SEDs, together with the COS spectra, we perform SED fitting and obtain \ovi cooling rates nearly identically to our procedures explained in Papers I and II.  As a brief overview, during SED fitting, we mask the emission lines and model the stellar continuum with \textsc{Prospector} \citep{2021ApJS..254...22J}, using a custom-built stellar template library with high resolution in the far-UV.  From this step, we obtain SFRs averaged over the past 10 Myr and 100 Myr.  For details on the SED fitting, we refer the reader to Paper I.  The only notable difference from the techniques previously reported is that we allow the normalization of the background spectrum to vary independently between the two COS detector chips, as opposed to having a single background normalization for the full \textit{FUSE} spectrum.

The stellar continuum models are then subtracted and the residuals are used to model the \ovi emission, accounting for attenuation effects from dust, molecular H$_2$, and metal absorption, both along the line of sight and intrinsic to the system itself.  \ovi luminosities are converted to cooling rates using \textsc{Cloudy} \citep{1998PASP..110..761F, 2023RMxAA..59..327C} simulations of cooling gas with appropriate physical conditions.  It is important to note that this conversion is model-dependent, and therefore \mdotovi strictly represents the cooling rate under the assumption of a single-pass steady cooling flow, with the adopted ionization and thermal models, rather than an actual measurement of a net deposition rate.  Including the effects of turbulence and mixing, with gas repeatedly crossing the \ovi emission peak \citep[as is seen in hydrodynamic simulations;][]{2017MNRAS.466..677G} would potentially dramatically change the inferred cooling rates.  We consider these effects and their implications in $\S$\ref{sec:discuss}.  For further details on the \ovi fitting and \textsc{Cloudy} modeling, we refer the reader to Paper II.  The elemental abundances that we use in our \textsc{Cloudy} models are obtained from \citet{2004AnA...420..135T} for Abell 1835.  CHIPS 1911, MACS 1931, and RX 1532 use the average cluster abundances measured in \citet{2017AnA...603A..80M}.  

We also note that the Phoenix Cluster is treated as a special case in our \textsc{Cloudy} models.  Since it uniquely hosts a powerful radiatively active quasar, AGN photoionization cannot be neglected when accounting for sources of \ovi ionization (as it is with the remainder of our systems).  Therefore, for Phoenix, we use the models presented in \citet{2025Natur.638..360R}, which simulate a parcel of cooling gas that is simultaneously irradiated by a luminous quasar at varying distances.


Additionally, we gather two measures of the ``maximal'' cooling rate derived from the X-ray temperature and luminosity, $\mdotmax = (2/5)(\mu m_p/kT)L_X$.  One is integrated out to the cooling radius $r_{\rm cool}$, defined as the point where the cooling time equals 3 Gyr \citep[following ][]{2018ApJ...858...45M}.  The other is integrated within the $2.5''$-diameter COS aperture, which we label \mdotmaxap.  These are calculated identically to the techniques used in Paper II, which use standard X-ray surface brightness modeling techniques on archival \textit{Chandra}-ACIS data.  For more details on this, we additionally refer the reader to \citet{2006ApJ...640..691V} and \citet{1997MNRAS.292..419W}.  It is worth noting that both \mdotmax and \mdotmaxap are important comparison points, despite all of our other cooling rates being strictly measured within the COS aperture.  Cooling X-ray gas may condense and flow inwards before emitting strongly in \ovi, making \mdotmax the more appropriate comparison, as opposed to the case where the cooling X-ray gas condenses mostly in-place, for which \mdotmaxap would be the better comparison \citep{2016A&A...595A.123M}.

Finally, we also gather spectroscopic X-ray cooling rate measurements from the literature, tracing the higher-ionization \ion{O}{7} and \ion{O}{8} (among others).  Standard (no absorption) models are obtained from \citet{2019MNRAS.485.1757L}, which we label \mdotx (their $\dot{M}_{\rm 1cie+1cf}$), and ``hidden'' (absorbed) cooling models are obtained from \citet{2022MNRAS.515.3336F, 2023MNRAS.521.1794F, 2023MNRAS.524..716F, 2024MNRAS.535.2173F, 2024MNRAS.535.2697I, 2025arXiv250814785F}, which we label \mdothcf.\footnote{Additionally, new \mdothcf measurements for RX 1532  have been obtained from A. C. Fabian 2026 (priv comm.), which include more data than the most recently published estimates, totaling $\sim 580$ ks, and further constraining the allowed solutions.}

\section{Results} \label{sec:results}

\subsection{Star formation rates}

\begin{figure}
    \centering
    \includegraphics[width=\columnwidth]{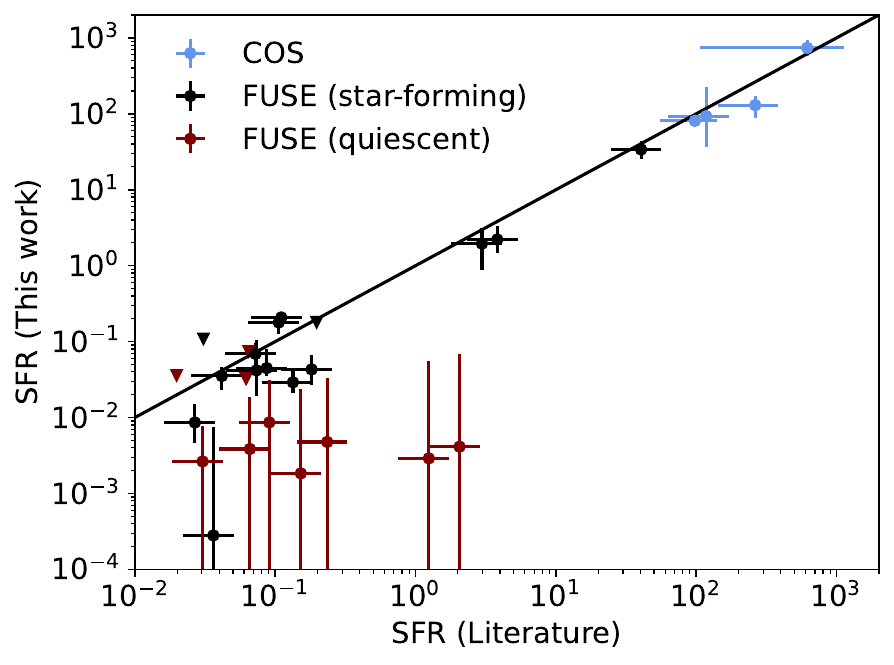}
    \caption{Our 100 Myr averaged SFRs are plotted as a function of literature-based SFRs. COS systems use SFRs from \citet{2018ApJ...858...45M} (which does not include CHIPS 1911), while \textit{FUSE} systems use the \textit{GALEX} FUV $+$ \textit{WISE} 24 ${\rm \mu m}$ scaling relation used in Paper I from \citet{2011ApJ...741..124H}. The $x$ errors for the \textit{FUSE} systems thus use the typical systematic error from the \citet{2018ApJ...858...45M} sample. The solid line shows 1:1 agreement.  \textit{FUSE} systems are split into star-forming and quiescent populations, based on the fraction of integrated UV light that comes from the young stellar population, with the cutoff at $f_{\rm young} = 0.03$. A clear quiescent population is present at low SFR, but disappears above $\sim 1\,\msunyr$.}
    \label{fig:sfr_m18}
\end{figure}

All 5 systems in our sample exhibit prominent SFRs on the order of $\sim 100~\msunyr$, as expected based on the nature of our sample selection.  SFRs averaged within the past 100 Myr agree well with the SFRs from the literature, catalogued in \citet{2018ApJ...858...45M}, as we show in Figure \ref{fig:sfr_m18}.  In paper I, we reported an offset between our SED-based SFRs and standard 2-band \textit{GALEX}+\textit{WISE} photometric SFRs in systems dominated by the UV upturn (i.e. old, UV-emitting stellar populations, shown in red).  This effect is absent from the COS systems, because their UV emission is dominated by truly young stellar populations.  Our modeled SFRs also exhibit lower uncertainties than the literature-based measurements, in all cases except for Abell 1835, where they are comparable.  This may simply reflect a difference in accounting, as \citet{2018ApJ...858...45M} average measurements from many different techniques, which will inherently account for systematic uncertainties.  Meanwhile, our uncertainties reflect a marginalization over all free parameters in our SED model, but this model is not flexible over all possible dimensions (for example, we assume a fixed Salpeter IMF, a parametric double exponential shape for the star formation history, a Calzetti dust extinction law, etc.).

\begin{figure}
    \centering
    \includegraphics[width=\columnwidth]{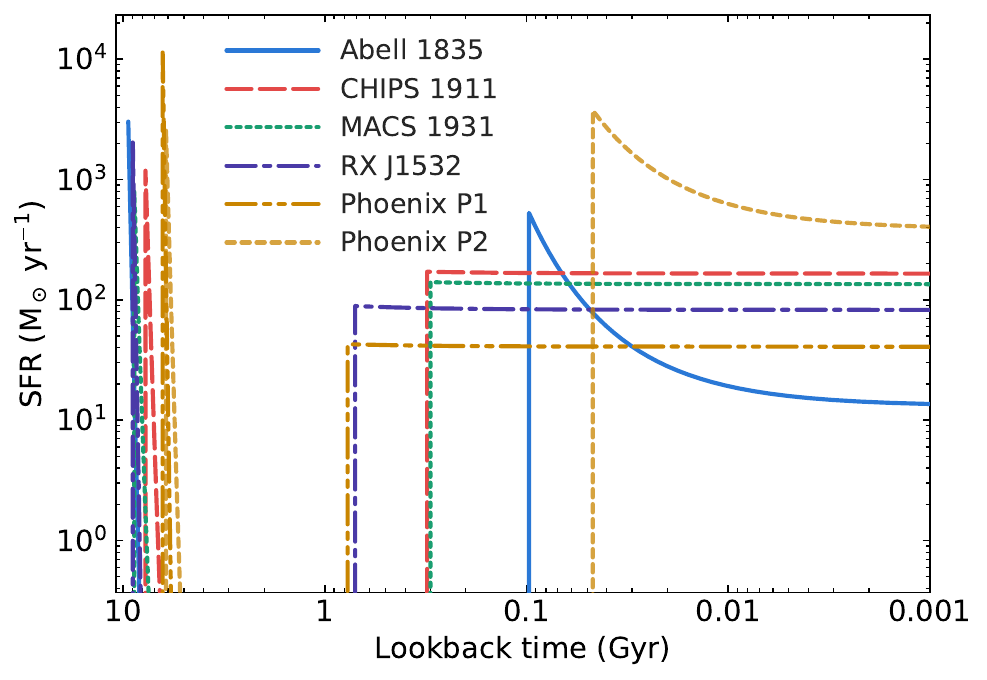}
    \caption{Parametric star formation histories are shown for all 5 systems in our sample.  Lookback time increases to the left, with large narrow spikes representing the old stellar populations, and young stellar populations becoming important $< 1$ Gyr and remaining so until the present (at the right edge of the plot).  Note the logarithmic scales of both axes.  No system reaches a present-day SFR below $10~\msunyr$.}
    \label{fig:sfh}
\end{figure}

The parametric double-exponential star formation histories for all 5 systems in our sample are shown in Figure \ref{fig:sfh}.  We observe unanimously that the preferred shape of the young stellar population is an extended tail of continued ongoing star formation over the past 0.1--1 Gyr, with no systems dropping below an instantaneous SFR of $10~\msunyr$.  While the exact shapes of these SFHs should not be interpreted too literally, this provides substantial evidence of ongoing star formation, rather than post-starburst activity, in these systems. One possible exception may be Abell 1835, where the presently-observed star formation rate is estimated to be nearly two orders of magnitude below the peak star formation rate, measured to have occurred 100 Myr ago. This tentative evidence for post-starburst activity is intriguing and will be examined further in a follow-up study using combined UV and optical spectroscopy.

\subsection{\ovi fluxes and kinematics}

\begin{deluxetable*}{llllllllllll}
\def\arraystretch{1.15}
\tabletypesize{\footnotesize}
\tablecaption{All \ovi fluxes and kinematics, along with relevant attenuation parameters.}
\label{tab:ovi}
\tablehead{
    \colhead{Name} & \colhead{$E(B-V)_{\rm gal}$} & \colhead{$E(B-V)_{\rm neb}$} & \colhead{$E(B-V)_{\rm diff}$} & \colhead{H$_{2,\rm gal}$} & \colhead{H$_{2,\rm int}$} & \colhead{Det?\tablenotemark{$*$}} & \colhead{$F_{\rm O\,VI,c1}/10^{-16}$\tablenotemark{$\dagger$}} & \colhead{$F_{\rm O\,VI,c2}/10^{-16}$\tablenotemark{$\ddagger$}} & \colhead{$v_{\rm los}$} & \colhead{FWHM} & \colhead{$D$\tablenotemark{$\S$}} \\
    \colhead{} & \colhead{mag} & \colhead{mag} & \colhead{mag} & \colhead{$\log({\rm cm^{-2}})$} & \colhead{$\log({\rm cm^{-2}})$} & \colhead{} & \colhead{$\ergscm$} & \colhead{$\ergscm$}& \colhead{\kms} & \colhead{\kms} & \colhead{Mpc}
}
\startdata
\hline
Abell 1835 & $0.026_{-0.0}^{+0.0}$ & $0.9_{-0.3}^{+0.3}$ & $0.0_{-0.008}^{+0.008}$ & $17_{-1}^{+2}$ & $15.3_{-0.2}^{+0.2}$ & Y & $4.1_{-0.5-0.6}^{+0.5+0.6}$ & $5.3_{-0.7-1.4}^{+0.7+1.4}$ & $50_{-40}^{+40}$ & $370_{-60}^{+60}$ & $1278$\\ 
CHIPS 1911+4455 & $0.065_{-0.001}^{+0.001}$ & $1.1_{-0.3}^{+0.2}$ & $0.0_{-0.008}^{+0.007}$ & $18_{-2}^{+2}$ & $15.6_{-0.4}^{+0.4}$ & N & $<6$ & $<7$ & $-$ & $-$ & $2760$\\ 
MACS J1931.8-2634 & $0.095_{-0.001}^{+0.001}$ & $1.0_{-0.3}^{+0.1}$ & $0.0_{-0.008}^{+0.008}$ & $17_{-2}^{+2}$ & $15.7_{-0.5}^{+0.5}$ & N & $<4$ & $<4$ & $-$ & $-$ & $1884$\\ 
RX J1532.9+3021 & $0.026_{-0.001}^{+0.001}$ & $1.53_{-0.07}^{+0.08}$ & $0.0_{-0.006}^{+0.005}$ & $18_{-2}^{+2}$ & $16.0_{-0.6}^{+0.6}$ & P & $1.1_{-0.3-1.2}^{+0.4+1.2}$ & $1.4_{-0.4-1.2}^{+0.5+1.2}$ & $-200_{-100}^{+400}$ & Unconstrained & $1947$\\ 
Phoenix SE & $0.015_{-0.001}^{+0.001}$ & $1.26_{-0.09}^{+0.07}$ & $0.0_{-0.006}^{+0.005}$ & $18_{-2}^{+2}$ & $15.17_{-0.09}^{+0.08}$ & Y & $29_{-1-3}^{+1+3}$ & $37_{-1-6}^{+1+6}$ & $200_{-20}^{+20}$ & $700_{-30}^{+20}$ & $3555$\\ 
Phoenix NW & $0.015_{-0.001}^{+0.001}$ & $0.3_{-0.2}^{+0.1}$ & $0.0_{-0.004}^{+0.005}$ & $17_{-1}^{+2}$ & $15.4_{-0.2}^{+0.2}$ & Y & $23_{-1-2}^{+1+2}$ & $30_{-1-4}^{+1+5}$ & $-70_{-20}^{+20}$ & $720_{-50}^{+60}$ & $3555$\\ 
\hline
\enddata
\tablenotetext{}{All reported \ovi fluxes are for the brighter $\lambda$1032\angstrom line.  The dimmer $\lambda$1038\angstrom line is enforced to be half of this flux in our model.  The first pair of uncertainties is the statistical uncertainty $\sigma_{\rm stat}$, while the second pair is the systematic uncertainty $\sigma_{\rm syst}$.}
\tablenotetext{}{All measurements are made within the COS PSA aperture, which is a circle with diameter $2.5''$.}
\tablenotetext{*}{Whether or not \ovi is detected.  ``Y'' means a strong detection exceeding both $2.5\sigma_{\rm stat}$ and $2.5\sigma_{\rm syst}$.  ``P'' means a possible detection exceeding $2.5\sigma_{\rm stat}$ but not $2.5\sigma_{\rm syst}$.  ``N'' means a non-detection.}
\tablenotetext{\dagger}{The flux after correcting for galactic metal absorption, galactic H$_2$ absorption, intrinsic H$_2$ absorption, and galactic dust extinction.}
\tablenotetext{\ddagger}{The flux after correcting for galactic metal absorption, galactic H$_2$ absorption, intrinsic H$_2$ absorption, galactic dust extinction, \textit{and} intrinsic dust extinction.}
\tablenotetext{\S}{The luminosity distance calculated using our standard cosmology.}
\end{deluxetable*}

We report fluxes and kinematics of the \ovi $\lambda 1032\angstrom$ line, as well as ancillary information, in Table \ref{tab:ovi}, mirroring Table 2 in Paper II.  $F_{\rm O\,VI,c1}$ is the flux after correcting for attenuation from galactic dust extinction, galactic metal absorption, and galactic and intrinsic H$_2$ absorption.  $F_{\rm O\,VI,c2}$, meanwhile, applies an estimated correction for intrinsic dust extinction, on top of all of the corrections applied to $F_{\rm O\,VI,c1}$.  As explained in Paper II, these two estimates of the flux are reported separately because the intrinsic dust correction that should be applied to gas in the $\sim 10^{5.5}\,{\rm K}$ phase is highly uncertain, with our correction relying on toy models of the phase structure in the gas, as well as the details of the dust distribution.

Following Paper II, we report \ovi measurements which exceed both $2.5\sigma_{\rm stat}$ \textit{and} $2.5\sigma_{\rm syst}$ as ``strong'' detections, those that exceed $2.5\sigma_{\rm stat}$ but not $2.5\sigma_{\rm syst}$ as ``possible'' detections, and all others as non-detections.  Following this definition, Abell 1835 and Phoenix are exceptionally strong detections (both exceeding $6\sigma_{\rm syst}$), while RX 1532 is a possible detection (barely reaching $3\sigma_{\rm stat}$, and $<1\sigma_{\rm syst}$).  In CHIPS 1911, a broad feature is observed at the location of \ovi, but this feature is only significant at the $\sim 2\sigma_{\rm stat}$ level, whereas no significant emission is observed in MACS 1931.  Therefore, for these two systems, we report $3\sigma$ upper limits.  

Notably, \citet{2017ApJ...835..216D} originally reported a non-detection of \ovi in the RX 1532 data.  They fit the continuum with a power law, including line features for Ly$\alpha$ (emission), Ly$\beta$ (absorption), and \ovi, and reported that the best-fit model for \ovi preferred absorption over emission.  While our results still only suggest a weak possibility of a detection, it is nevertheless interesting that they differ. We believe this is most likely attributable to a difference in model continuum shapes near Ly$\beta$---our stellar population models predict very wide, asymmetric, and complex absorption troughs near Ly$\beta$ (see the bottom-right panel of Figure \ref{fig:cos_apertures}), which cannot be reproduced with the simple treatment used by \citet{2017ApJ...835..216D}.  Our recovered flux falls within their reported $3\sigma$ upper limit of $2.34 \times 10^{-16}~\ergscm$.  Additionally, we note that the line width in RX 1532 is unconstrained, nominally preferring a very broad feature, which artificially inflates the flux and may result in a false positive. However, we still recover a weak $3\sigma$ detection even when the line width is capped at $720~\kms$ (the maximum width observed in any other system, Phoenix), and thus report the line as a detection rather than an upper limit.

Our recovered fluxes for the Phoenix Cluster are lower than reported in \citet{2015ApJ...811..111M} by roughly an order of magnitude, due to a difference in our prescription of the intrinsic dust extinction correction.  Our correction is more conservative by assuming a less-than-unity dust covering fraction in the $10^{5.5}~{\rm K}$ gas phase, which leads to marginal, order-unity correction factors on the final line fluxes (for details, see Paper II).

Abell 1835 exhibits similar kinematics as the parent \textit{FUSE} sample, whereas Phoenix is significantly broadened, with line full-widths at half-maximum (FWHMs) of $\gtrsim 700~\kms$.  In all cases, the line widths are far greater than COS's line spread function ($\sim 15-20~\kms$ for the medium-resolution gratings, and $\sim 100-200~\kms$ for the low resolution gratings, relevant only for RX 1532).  Broad \ovi may trace hot phase turbulence, unresolved multiphase cloud motions, bulk flows, and/or feedback-driven circulation.  It is difficult to distinguish these sources from our aperture-integrated \ovi measurements alone, but inferences may be made by comparing the kinematics to other gas phases, for example with H$\alpha$ or CO.  We leave a more detailed analysis of the kinematics of multiphase gas in these systems to future work.

Also noteworthy is a prominent emission feature at a rest wavelength of $1034.2\,\angstrom$, between the two \ovi lines, in Abell 1835's spectrum.  This wavelength does not correspond to any prominent emission lines, either in the rest frame or the observed frame.  We therefore believe that this corresponds to a high-velocity component of the \ovi $\lambda$1032 line, redshifted to $+650\,\kms$ relative to the rest frame, arising from an inflow or outflow of hot gas.  This may be related to the cold outflow of molecular gas which has been observed in this system's CO emission with ALMA \citep{2014ApJ...785...44M}.  No similar redshifted emission can be observed from the \ovi $\lambda$1038 line because of an extremely bright triplet of airglow lines: \ion{O}{1} $\lambda$$\lambda$1302.17,1304.86,1306.03\,\angstrom.  Fits to the \ovi lines for each system are shown in the Appendix.

\subsection{Cooling rates}

\begin{deluxetable*}{lllllllllll}
\def\arraystretch{1.15}
\tabletypesize{\footnotesize}
\tablecaption{All cooling rate measurements and upper limits.}
\label{tab:cooling_rates}
\tablehead{
    \colhead{Name} & \colhead{SFR$_{100}$} & \colhead{SFR$_{10}$} & \colhead{$\dot{M}_{\rm O\,VI,cp}$} & \colhead{$\dot{M}_{\rm O\,VI,cd}$} & \colhead{$\dot{M}_{\rm O\,VI,comb}$} & \colhead{$\dot{M}_{\rm O\,VI,snc}^*$} & \colhead{\mdotx} & \colhead{$\dot{M}_{\rm X,spec,hcf}$} & \colhead{\mdotmaxap} & \colhead{\mdotmax}
}
\startdata
\hline
Abell 1835 & $90_{-60}^{+130}$ & $14_{-8}^{+16}$ & $210_{-30-90}^{+30+80}$ & $350_{-40-160}^{+40+140}$ & $210_{-30-90}^{+30+80}$ & $170_{-30-80}^{+30+80}$ & $80_{-30}^{+30}$ & $>70$ & $18_{-4}^{+4}$ & $800_{-300}^{+300}$\\ 
CHIPS 1911+4455 & $160_{-70}^{+90}$ & $145_{-60}^{+82}$ & $<600$ & $<1000$ & $<700$ & $<700$ & $-$ & $-$ & $72_{-18}^{+18}$ & $700_{-200}^{+300}$\\ 
MACS J1931.8-2634 & $130_{-40}^{+40}$ & $113_{-35}^{+32}$ & $<200$ & $<300$ & $<200$ & $<200$ & $160_{-70}^{+40}$ & $>700$ & $138_{-34}^{+34}$ & $900_{-300}^{+200}$\\ 
RX J1532.9+3021 & $80_{-10}^{+10}$ & $73_{-12}^{+13}$ & $60_{-20-60}^{+20+60}$ & $110_{-30-100}^{+40+100}$ & $70_{-20-60}^{+20+60}$ & $50_{-20-60}^{+20+60}$ & $<300$ & $700_{-100}^{+200}$ & $136_{-33}^{+33}$ & $1200_{-300}^{+300}$\\ 
Phoenix & $740_{-170}^{+200}$ & $503_{-128}^{+120}$ & $4200_{-200-2800}^{+200+9000}$ & $6500_{-200-4600}^{+200+14500}$ & $4200_{-200-2800}^{+300+9000}$ & $4200_{-200-2800}^{+300+9000}$ & $400_{-200}^{+200}$ & $>800$ & $743_{-183}^{+183}$ & $3000_{-1100}^{+700}$\\ 
\hline
\enddata
\tablenotetext{}{All measurements are given in units of \msunyr.}
\tablenotetext{}{For \mdotovi measurements, the first pair of uncertainties is the statistical uncertainty $\sigma_{\rm stat}$, while the second pair is the systematic uncertainty $\sigma_{\rm syst}$.}
\tablenotetext{*}{$\dot{M}_{\rm O\,VI,snc}$ include a correction term for \ovi emission from supernova remnants.  See Paper II for details.}
\end{deluxetable*}

We report all cooling rate measurements, including star formation rates, in Table \ref{tab:cooling_rates}, mirroring Table 3 in Paper II.  Note $\dot{M}_{\rm O\,VI,cp}$ and $\dot{M}_{\rm O\,VI,cd}$ refer to purely isobaric (constant pressure) and purely isochoric (constant density) cooling rates, respectively, while $\dot{M}_{\rm O\,VI,comb}$ refers to a combined isobaric-isochoric model.  $\dot{M}_{\rm O\,VI,snc}$ includes an additional correction for \ovi contamination from supernovae (SNe), as explained in Paper II.  In most cases, this represents a small fraction of the total observed \ovi luminosity, with $L_{\rm O\,VI,SNe}$ of order $10^{40}$ \ergs for these systems.  Both pointings of the Phoenix Cluster have been combined in this table to provide a single integrated cooling rate measurement.  

The table lists both statistical and systematic uncertainties for all \ovi cooling rates, the procedure for which is also explained in Paper II.  We note here, however, another exception regarding the Phoenix Cluster:  Due to the more complex nature of these calculations, which incorporate AGN photoionization, additional systematic uncertainties have been included which account for the uncertainty in the absolute bolometric luminosity of the AGN.

\section{Discussion} \label{sec:discuss}

With the new cooling rate measurements in hand, we provide an updated look at the correlations between cooling rates presented in Paper II.  These correlations are modeled as simple power laws, with a line in log-log space $y = ax + b \pm \sigma$, parametrized symmetrically in $y$ and $x$ by the angle that satisfies $a = \tan\theta$, perpendicular intercept $b = b_\perp/\cos\theta$, and perpendicular scatter $\sigma = \sigma_\perp/\cos\theta$.  The likelihood uses orthogonal displacements, and accounts properly for upper and lower limits.  This methodology all follows precisely from our procedures in Paper II, and from \citet{2010arXiv1008.4686H}.  

The one area that the current analysis differs is in how we model the \sfr-\mdotmax relation.  Previous works have observed significant non-linearity in this relation, with a floor at $\sfr \sim 0.1\,\msunyr$ \citep{2018ApJ...858...45M}, reached at $\mdotmax \lesssim 10~\msunyr$ \citep{2022ApJ...940..140C}.  In these studies, it is unclear if this is a sensitivity floor in the \sfr measurements, or whether it reflects a true physical floor driven by some baseline source of star formation uniformly present across the probed mass range.  Therefore, in this study, we are motivated to model the \sfr-\mdotmax relation with a floor, and compare it to the floor seen in previous studies.  This is a simple change to the model, $y = \max(ax+b, c)$, but this breaks the symmetry in $x$ and $y$ and thus precludes us from performing an orthogonal regression for this relation.  All other relations are fit orthogonally.

\subsection{\mdotovi vs. \mdotmax and \mdotmaxap}

\begin{figure*}[htb]
    \centering
    \includegraphics[width=0.7\linewidth]{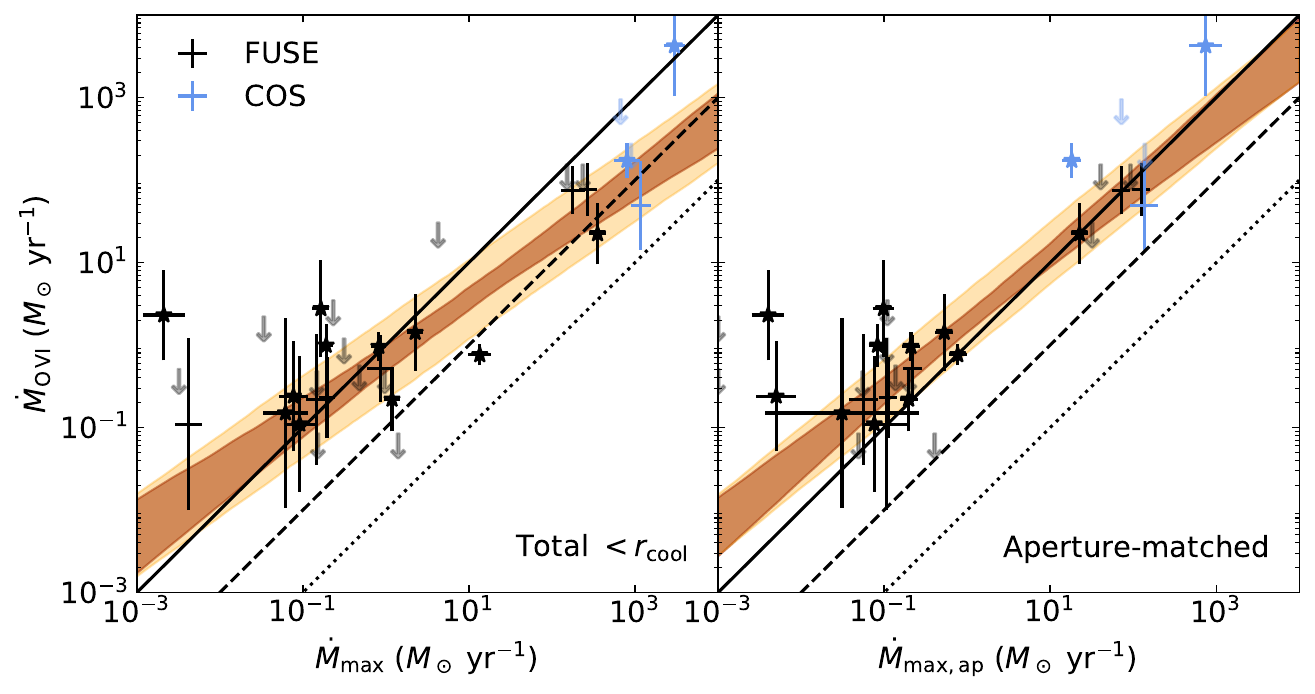}
    \caption{We show \ovi-based cooling rates ($\dot{M}_{\rm O\,VI,snc}$) against the maximal cooling rate integrated within $r_{\rm cool}$ (\mdotmax; left panel) and aperture-matched to the \textit{FUSE} and COS data (\mdotmaxap; right panel).   The 29 systems from the \textit{FUSE} study are shown in black, while the 5 new COS systems are shown in blue. The error bars show the combined statistical and systematic errors.  Strong detections (``Y'' in Table \ref{tab:ovi}) are marked with stars, while marginal detections (``P'' in Table \ref{tab:ovi}) have no marker.  Translucent arrows, with colors matching \textit{FUSE} or COS, denote upper limits in either $x$ or $y$. The solid, dashed, and dotted lines show $y=x$, $y=0.1x$, and $y=0.01x$, respectively.  The yellow shaded region shows the $1\sigma$ intrinsic scatter region around the line of best fit, while the red shaded region shows the $1\sigma$ region of measurement uncertainty around the best fit, both obtained from bootstrapping a linear model with a likelihood that includes both detections and upper limits (as explained in Paper II). The best fits are $\log\mdotovi = (0.71_{-0.10}^{+0.09})\log\mdotmax - 0.17_{-0.15}^{+0.14} \pm 0.42_{-0.21}^{+0.12}$ and $\log\mdotovi = (0.81_{-0.08}^{+0.10})\log\mdotmaxap + 0.21_{-0.10}^{+0.10} \pm 0.37_{-0.37}^{+0.11}$.}
    \label{fig:movi_mmax}
\end{figure*}

\mdotovi is shown as a function of \mdotmax and \mdotmaxap in Figure \ref{fig:movi_mmax}.  For the most part, it is observed that the COS systems continue the trend seen in the \textit{FUSE} systems, where the low-mass groups and isolated ellipticals appear to have an enhanced cooling efficiency (\mdotovi/\mdotmax), while the high-mass clusters' efficiency is reduced.  The trend is seen in both the fully integrated and the aperture-matched versions of the relation, though the new COS measurements serve to reduce its significance, being distinct from a 1:1 relationship at $\sim 3\sigma$, as opposed to $\sim 4\sigma$ if only the \textit{FUSE} data is used.

In Paper II, we argued that this trend might be explained by differences in the radial distribution and confinement of feedback from AGN and SNe between groups and clusters.  The Intragroup Medium (IGrM) is higher in entropy ($\propto n_e^{-2/3}kT$), and therefore lower in pressure ($\propto n_ekT$), than expected based purely on self-similar scaling arguments from clusters \citep{2003ApJ...593..272V}.  The AGN jets can thus act on larger physical scales in groups than in clusters, having encountered less ram pressure, and leading to a deposition of energy at larger radii, outside the core regions.  According to \citet{2020ApJ...899...70V}'s ``black hole valve'' mechanism, the combination of this larger-scale feedback and SNe-driven outflows can create a ``stagnation radius'', above which gas may be easily ejected from the system, and below which it cannot (instead cooling and condensing onto the black hole).  Meanwhile, at the cluster scale, the higher pressure ICM confines the AGN jets and SNe-driven outflows, which mostly closes the feedback loop and establishes a global thermodynamic equilibrium.  Remaining cooling may then develop at the 1--10\% level from local instabilities \citep{2012ApJ...746...94G, 2015Natur.519..203V, 2026arXiv260527504B, 2026arXiv260527508P, 2026arXiv260527511C}.  We note that this is merely one possible interpretation of the data, while the observed trends could also be explained by other factors, such as the AGN duty cycle or \ovi emissivity.  Even at a fixed global cooling power, jet-driven stirring and uplift can redistribute where coronal gas appears, depending on the local ratio of the cooling time to the eddy time ($\mathcal{C} = t_{\rm cool}/t_{\rm eddy}$), with $\mathcal{C} \sim 1$ identifying the condensation-prone regime \citep{2018ApJ...854..167G, 2026arXiv260527504B}.  Therefore, distinguishing these mechanisms further will require spatially resolved coronal line measurements, which are unavailable and largely unobtainable with current instrumentation.  

The new COS data largely align with the trends seen in the \textit{FUSE} data, with the new detections and upper limits falling adjacent to the highest-mass \textit{FUSE} clusters.  The obvious exception is the Phoenix Cluster, notorious for its extreme starburst and high \sfr/\mdotmax \citep{2012Natur.488..349M}, which lies close to the 100\% line in \mdotovi/\mdotmax. Although this measurement suffers from the most systematic uncertainty, it nevertheless supports the picture that this cluster may be experiencing the closest thing to an uninhibited cooling flow that we know of, and it resolves the tension in \citet{2015ApJ...811..111M} where the measured $\mdotovi$ exceeded $> 10\mdotmax$.  Another interesting exception is Abell 1835 in the \mdotmaxap case, which may hint at an extended origin for the hot gas feeding into the core, an idea which has also been discussed from its X-ray spectroscopy \citep{2016A&A...595A.123M}.

\subsection{\mdotovi vs. \mdotx and \mdothcf} \label{sec:mdotovi_mdotx}

\begin{figure*}
    \centering
    \includegraphics[width=0.7\linewidth]{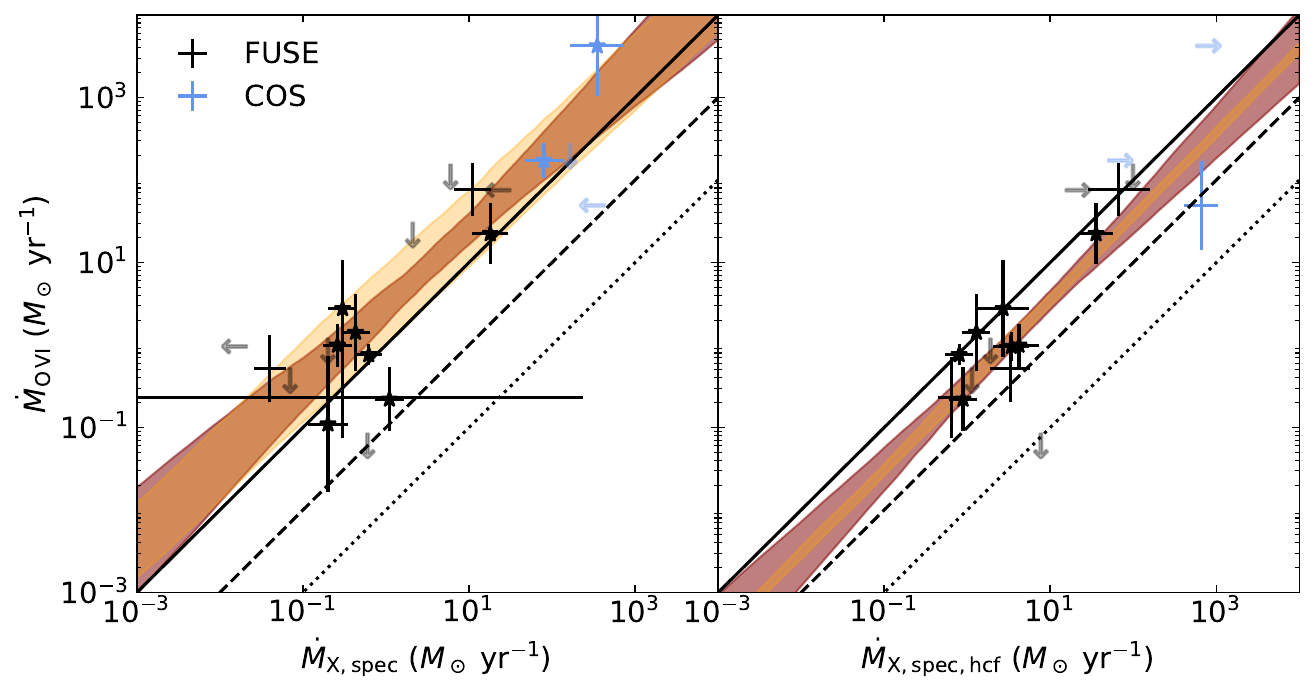}
    \caption{As Figure \ref{fig:movi_mmax}, but showing $\dot{M}_{\rm O\,VI,comb}$ as a function of \mdotx (left) and \mdothcf (right).  The supernovae-corrected values $\dot{M}_{\rm O\,VI,snc}$ are \textit{not} used here, as the X-ray spectroscopic cooling rates would also require similar corrections. We therefore opt to show the correlation between the uncorrected values rather than applying uncertain corrections to both axes and potentially introducing spurious correlations (since both corrections would scale with halo mass and \sfr).  The best fits, shown in red (covering the measurement uncertainty) and yellow (covering the system-to-system scatter) are $\log\mdotovi = (0.90_{-0.16}^{+0.20})\log\mdotx + 0.46_{-0.22}^{+0.20} \pm 0.57_{-0.36}^{+0.09}$ and $\log\mdotovi = (0.90_{-0.08}^{+0.21})\log\mdothcf - 0.45_{-0.19}^{+0.11} \pm 0.07_{-0.07}^{+0.02}$.}
    \label{fig:movi_mx}
\end{figure*}

\mdotovi is shown as a function of \mdotx and \mdothcf in Figure \ref{fig:movi_mx}.  Relative to \mdotx, most of the \textit{FUSE} systems exhibit enhanced \mdotovi by a factor of $\sim 3$ (with large scatter).  Excluding Phoenix, the COS data at the high-mass end appear more consistent with 1:1, although 2/3 of these measurements are upper limits, making it difficult to tell how significant this difference is.  Relative to \mdothcf, the \textit{FUSE} systems have suppressed \mdotovi by a factor of $\sim 3$ (with a smaller scatter), with the COS systems essentially continuing this trend.

Both of these relations show no strong preference for a non-unity slope.  In paper II, we therefore concluded based on the intercepts that either (1) X-ray absorption plays an important role, but not at 100\% covering fraction (as assumed by the \mdothcf models), placing the truth somewhere in between \mdotx and \mdothcf; or (2) \mdotovi is enhanced and \mdotx is suppressed due to turbulent mixing, with absorption playing a minimal role.

Here, the addition of the COS systems leaves these interpretations as the leading hypotheses, essentially unchanged.  The only potential deviation depends on how one interprets the 3 COS measurements which appear consistent with $\mdotovi/\mdotx \sim 1$, within the scatter.  If true, the high-mass end of this relation would require neither of the above explanations, being more suggestive of steady, unabosrbed cooling at $\sim 10-20\%$ of the maximal rate \mdotmax.  Ignoring Phoenix, this favors a sub-linear trend in \mdotovi/\mdotx that may be indicative of a smooth transition from mixing-dominated systems at low \mdotx to cooling-dominated systems at high \mdotx.  We explore this possibility further in the next few subsections, with the additional context of the SFRs.

\subsection{\sfr vs. \mdotmax} \label{sec:sfr_mdotmax}

\begin{figure*}
    \centering
    \includegraphics[width=\linewidth]{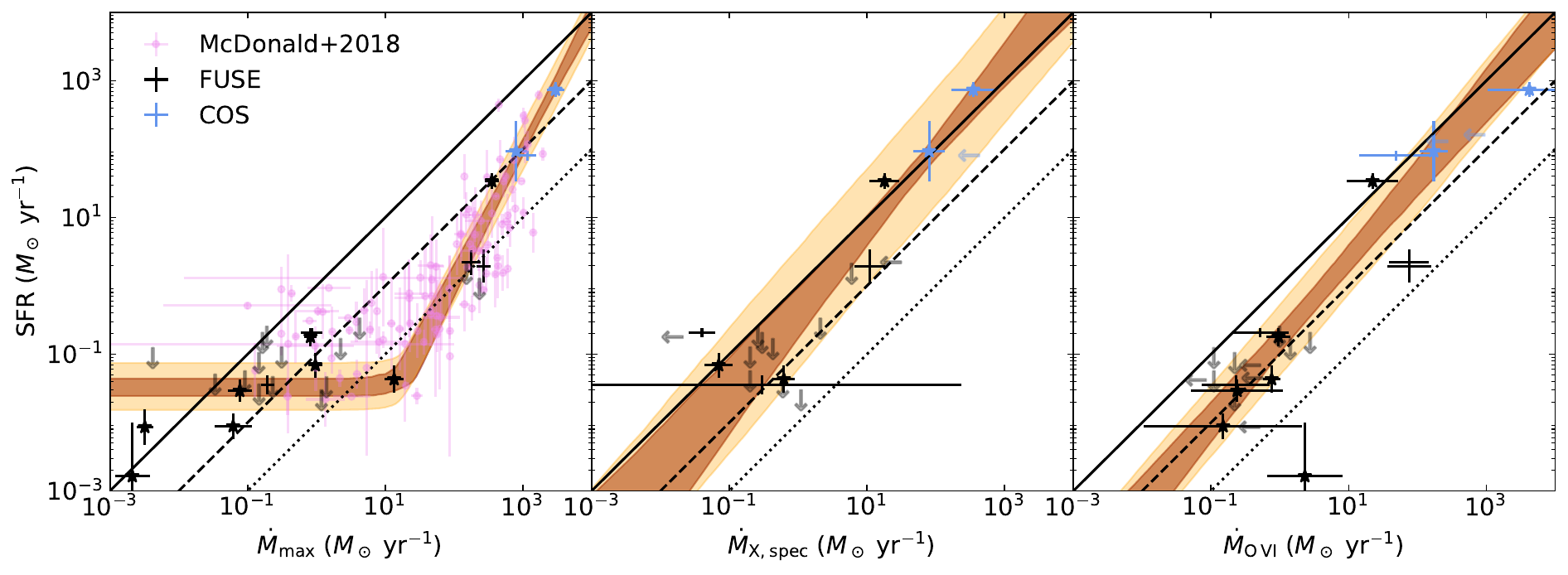}
    \caption{As Figure \ref{fig:movi_mmax}, but showing \sfr as a function of \mdotmax, \mdotx, and $\dot{M}_{\rm O\,VI,snc}$. All panels use 100 Myr averaged SFRs. The leftmost panel applies our constant floor model, while the two right panels do not impose constant floors.  The left panel also shows data from \citet{2018ApJ...858...45M} in violet for comparison purposes.  The best fits are $\log\sfr_{100} = \max((1.94_{-0.17}^{+0.22})\log\mdotmax - 3.80_{-0.69}^{+0.56}, -1.48_{-0.13}^{+0.12}) \pm 0.37_{-0.13}^{+0.04}$; $\log\sfr_{100} = (1.18_{-0.20}^{+0.23})\log\mdotx - 0.33_{-0.43}^{+0.31} \pm 0.96_{-0.38}^{+0.23}$; and $\log\sfr_{100} = (1.28_{-0.15}^{+0.19})\log\mdotovi - 0.85_{-0.21}^{+0.21} \pm 0.60_{-0.26}^{+0.11}$.}
    \label{fig:sfr_movi}
\end{figure*}

\sfr is shown as a function of \mdotmax, \mdotx, and \mdotovi in Figure \ref{fig:sfr_movi}.  In all panels, we use the 100 Myr averaged SFRs.

First focusing on the \sfr-\mdotmax relation, the new COS data all lie near $\sim 10\%$, but sit above the highest point in the \textit{FUSE} sample (Perseus), suggesting a far steeper trend at the high-mass end than we inferred previously (using a model with no floor).  However, the $\sfr/\mdotmax$ ratio in these systems is quite similar to those in the low-mass end, which also reach roughly $\sim 10\%$, due to the apparent floor in the relation below $\mdotmax \lesssim 10~\msunyr$.  This may be thought of as a consequence of both the trends in $\mdotovi/\mdotmax$ (decreasing with higher mass systems) and $\sfr/\mdotovi$ (increasing with higher mass systems), leading to a net neutral effect on $\sfr/\mdotmax$.  

Notably, the floor we measure from our sample sits at $\sfr \sim 0.03~\msunyr$, which is nearly 1 dex smaller than the floor in \citet{2018ApJ...858...45M}.  This suggests that the previously observed floor was not physical, but reflective of sensitivity limits.  At the same time, it may not be accurate to impose a floor on our sample either, as a few systems are observed well below $0.03~\msunyr$ (down to $10^{-3}$), and there may be a continued dependence on \mdotmax well below the ``knee'' of the relation at $\mdotmax \sim 10~\msunyr$.  For example, a purely linear power law with a slope of $\sim 1$ and an intercept of $\sim -1$ would fit the data nearly as well as the floor model, although it would over-predict data from $\mdotmax \sim 10-100~\msunyr$ in the \citet{2018ApJ...858...45M} sample by nearly an order of magnitude.  In both our sample and the \citet{2018ApJ...858...45M} samples, there is a real dearth of systems at high \sfr and $\mdotmax \sim 10-100~\msunyr$.  Given that we should be \textit{more} sensitive to these systems than the low-\sfr ones we actually observe in this range, this likely reflects a true lack of such systems in the underlying distribution, motivating the floor model.

\subsection{\sfr vs. \mdotx and \mdotovi}

Turning towards the \sfr-\mdotx and \sfr-\mdotovi relations (Figure \ref{fig:sfr_movi}), we note that there are 2 distinct regimes.  At low \sfr, the \sfr is $\sim 10\%$ of the spectroscopic cooling rate, whether we use X-ray or \ovi.  This is primarily driven by the \textit{FUSE} sample, and is consistent with what we observed in Paper II.  Whereas, at high \sfr, the \sfr reaches closer to $\sim 100\%$ of the spectroscopic cooling rate, also regardless of whether we use X-ray or \ovi.  The difference in regimes is driven by the new COS measurements (and Perseus), which stand in stark contrast to the \textit{FUSE} sample.  This observation implies that, in the systems with the highest cooling/star formation rates, the picture is quite simple: the X-ray spectroscopic cooling rates, UV spectroscopic cooling rates, and star formation rates all tend to converge on similar answers, suggesting that 10\% of the low-entropy gas is indeed cooling and forming stars.   System-to-system variation still remains, however, with a large scatter indicative of variable condensation and recycling and delayed star formation / reservoir consumption \citep{2017MNRAS.466..677G, 2026arXiv260527511C}.  Phoenix, being a clear outlier, is the cleanest example of this, with $\sfr/\mdotovi \sim 20\%$ and large systematic error bars.

These results are consistent with a few possible interpretations.  On one hand, there may be a smooth transition from $\sfr \sim 0.1\mdotovi$ at low \sfr to $\sfr \sim \mdotovi$ at high \sfr.  This implies that there is a changing balance between the different physical processes that generate \ovi (and higher-ionization X-ray lines), with the high-\sfr systems being more weighted towards net radiative cooling, and the low-\sfr systems containing (relatively) more emission from turbulent mixing and thermal recycling \citep{1990MNRAS.244P..26B, 2010ApJ...719..523K}, while retaining a common feedback-regulated multiphase cycle.  In Paper II, we discussed turbulent mixing as one mechanism which could explain $\sfr/\mdotovi \sim 10\%$ in isolation for the \textit{FUSE} sample, which remains relevant here.  Alternatively, it may be that the \sfr/\mdotovi ratio is forced to converge towards 1:1 at high \sfr, simply due to a tighter response time between the cooling and the star formation.  The molecular gas depletion timescale goes as $t_{\rm dep} \sim M_{\rm H_2}/\sfr$, while its nominal replenishment timescale goes as $t_{\rm rep} \sim M_{\rm H_2}/\mdotovi$.  Therefore, given equal H$_2$ masses, the more star-forming, more cooling systems would also inherently have faster depletion and replenishment times.  In contrast, the more quiescent systems would have longer timescales which allow for more stochastic variations and lags to accumulate, drifting the \sfr and \mdotovi apart.  In reality, the H$_2$ mass also varies substantially across the systems in our sample, so a more rigorous test of this scenario would require an explicit calculation, using CO measurements (for example from ALMA) to directly compare timescales, which we leave to future work.

On the other hand, it is possible that sample selection bias is responsible for the observed trend.  That is, the physical slope may be shallower, but with significant scatter that is not properly captured by either the \textit{FUSE} or COS samples in isolation, with the \textit{FUSE} sample preferrentially missing high-SFR systems, and the COS sample missing low-SFR systems.  In the case of COS, systems have been specifically selected as targets based on their hosting extreme starbursts and massive cooling flows \citep[e.g.][]{2018ApJ...858...45M, 2021ApJ...907L..12S}, so a level of selection bias is expected. Notably, Abell 1795 and Abell 2597 from the \textit{FUSE} sample exhibit similar \mdotovi as Abell 1835 and RX 1532, but lie $>1$ dex lower in \sfr, while the Perseus cluster from the \textit{FUSE} sample, known for being an extreme cool core, is more similar to the COS systems than to anything else from the \textit{FUSE} sample. Additionally, intrinsic scatter is expected in the relation simply due to the stochasticity of cooling, but this may be further exacerbated if the measured \sfr, \mdotx, and \mdotovi probe cooling on different timescales.  

In the following sections, we go through these possibilities in more detail.

\subsection{The Role of Sample Selection Bias} \label{sec:selectionbias}

First, we consider the potential role of sample selection biases.  We measure the expected level of sample bias in the COS systems by comparing the observed occurrence rate of $\sfr_{100} \gtrsim 80~\msunyr$ BCGs in the COS sample to their occurrence rate in the high-mass CEREAL sample \citep{2026ApJ..1000..252W}, which is an unbiased low-$z$ cluster sample with a simple and well understood selection function ($0.15 < z < 0.25$; $4.5 < (M_{500}/10^{14}\,\msun) < 10.0$). In COS, we observe $5/5$ BCGs with $\sfr_{100} > 80~\msunyr$ (by construction), while in CEREAL, considering only the cool core clusters, the fraction is $1/14$.  Therefore, under a completely unbiased selection process, the probability of observing $5/5$ such systems would be $p = 1.86 \times 10^{-6}$, corresponding to a deviation of $4.6\sigma$.  

\begin{figure}
    \centering
    \includegraphics[width=0.48\textwidth]{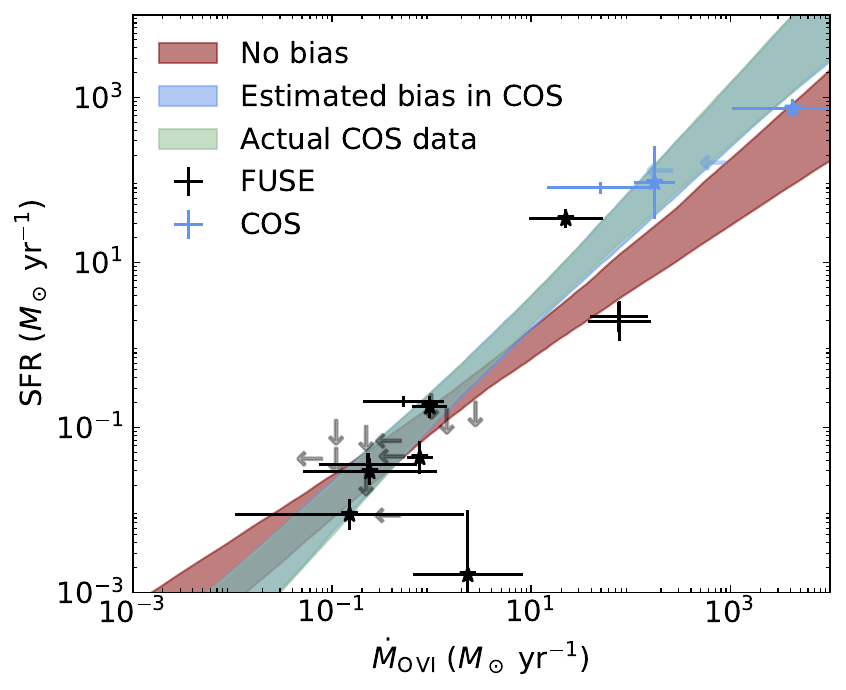}
    \caption{We test the results of our \sfr/\mdotovi relation against sample bias in the high-\mdotovi/\mdotmax COS systems.  In this Figure, SFRs are averaged over 100 Myr.  The red region shows the best fit for \textit{FUSE} data alone, with no bias.  The blue region shows the best fit to the mock COS observations with a cut of $80~\msunyr$.  And the green region shows the best fit to the real COS observations. The \sfr/\mdotovi relation is consistent with the biased fit.}
    \label{fig:selectionbias}
\end{figure}

Under the assumption that the \textit{FUSE} sample traces the true underlying \sfr/\mdotovi distribution, we measure $\sfr_{100}/\mdotovi = 10^{-0.94 \pm 0.64}$.  We then forward model mock COS observations which trace this distribution with the measured level of bias.  To do this, we take \mdotovi to be the same as the measured COS values, then apply the $\sfr/\mdotovi$ distribution to generate mock \sfr measurements, and apply a cut at $\sfr = 80~\msunyr$ (matching the observed threshold).  We then fit a linear power law to the mock observations, and compare it to the model that we observe from the real observations. The test results are shown in Figure \ref{fig:selectionbias}.

By construction, the nominal fit with no injected bias traces the \textit{FUSE} systems well, and essentially reproduces our best fitting models from Paper II.  In contrast, the observed fit is consistent with the biased model within $1\sigma$.  In other words, the observed \sfr/\mdotovi trend is roughly consistent with being explained purely by sample selection bias.  Regarding the other observed trends, we expect given the similarity between \mdotovi and \mdotx that a similar result holds for the \sfr/\mdotx relation.  However, the \sfr/\mdotmax relation is biased in both \sfr and \mdotmax in the same direction, so we expect a net neutral effect on the observed trend.

We note that this test is by no means definitive proof that the observed trends are or are not caused by sample selection bias.  It is an approximate calculation which merely shows that sample selection is a plausible explanation for the observed slope.  We have ignored any potential biases in the \textit{FUSE} sample itself, and in the observed distribution of \mdotovi, both of which are much more complicated than the \sfr distribution in COS.  Also, we emphasize here that the fact that the COS sample is biased towards strongly star-forming systems only affects our ability to understand the underlying population of cluster cores. It does not mean that our measurements themselves are biased and our conclusion that starburst BCGs are fueled directly by the cooling ICM stands.

\subsection{The Role of Turbulent Mixing}

We now consider the feasibility of turbulent mixing layers to explain a slope $> 1$ in the \sfr-\mdotovi relation.  If the physical conditions of turbulent mixing layers change with the ambient temperature (and thus halo mass), or if the relative contribution of \ovi emission from mixing layer regions and pure cooling regions changes with ambient temperature, this could explain a non-unity slope in \sfr-\mdotovi.  For example, low-mass groups may have relatively more \ovi emission from mixing, while high-mass clusters may have relatively less, which would cause the slope to steepen.

We test this hypothesis by solving a small set of analytic mixing layer solutions, from the same models we used in Paper II \citep{2023ApJ...950...91C}, at various hot phase temperatures: $T_{\rm hot} = 0.5, 1.0, 3.0$, and $6.0 \times 10^{7}$ K, and calculating the \ovi emission per unit mass, $\Gamma_{\rm ML}$, relative to that expected from a pure cooling flow model, $\Gamma_{\rm CF}$.  The behavior of the latter ($\Gamma_{\rm CF}$) with temperature is well known \citep[e.g.][]{1988ASIC..229...63C}, with a sharp rise at the temperatures most favorable to the ionization of the given line ($3 \times 10^5$ K for \ovi), followed by a flattening at higher temperatures.  The ranges probed here at $>5 \times 10^6$ K  are all well above the peak of \ovi ionization, so the behavior of $\Gamma_{\rm CF}$ should be essentially flat.

These models are idealized 1-dimensional steady-state plane-parallel analytic prescriptions, assuming a monotonic temperature profile and a constant mass flux.  They are not full 3D hydrodynamical simulations, and therefore they do not capture the phase structure of the gas around the mixing layer.  The specific value of $\Gamma_{\rm ML}$ is also sensitive to many parameters which are not constrained by our observations (such as the effective conductivity and viscosity).  Additionally, we caution that the upper bound of the parameter space on $T_{\rm hot}$ that we probe is an order of magnitude larger than the range which these models were tested and calibrated on in \citet{2023ApJ...950...91C}.  Due to these large uncertainties, we do not interpret the absolute value of $\Gamma_{\rm ML}$ literally, but instead focus on the trend in $\Gamma_{\rm ML}$ with $T_{\rm hot}$, while holding all other parameters fixed.

The analytic prescription, as outlined by \citet{2023ApJ...950...91C}, is defined by dimensionless variables for the width of the mixing layer $h$, Mach number of the relative shear velocity $\mathcal{M}_{\rm rel}$, conductivity strength $f_\nu$, conductivity floor $\kappa_0$, Prandtl number ${\rm Pr}$, and opacity $\tau$.  Therefore, keeping these variables fixed while modifying $T_{\rm hot}$ will create a set of self-similar models, rather than keeping the physical conditions themselves fixed.  In particular, $h$ and $\mathcal{M}_{\rm rel}$ both scale with temperature as $T_{\rm hot}^{-1/2}$, as their definitions are relative to the sound speed $c_{s,\rm hot} \propto T_{\rm hot}^{1/2}$.  As we wish to keep the physical conditions fixed to study the change in $\Gamma_{\rm ML}$ with $T_{\rm hot}$, our set of simulations has $h$ and $\mathcal{M}_{\rm rel}$ which vary according to $T_{\rm hot}^{-1/2}$.  Meanwhile, $f_\nu$, $\kappa_0$, ${\rm Pr}$, and $\tau$ are temperature-independent.  We keep $\kappa_0$ and $\tau$ fixed at their fiducial values, while setting $f_\nu$ and ${\rm Pr}$ based on the energy conservation condition \citep[$\mathcal{H}_{\rm visc}({\rm Pr}) \equiv \dot{m}\mathcal{M}_{\rm rel}^2/2$, where $\mathcal{H}_{\rm visc}$ is the viscous heating; see][for details]{2023ApJ...950...91C}.  Note that this does not mean that $\Gamma_{\rm ML}$ is independent of these parameters, but, as mentioned, we are primarily interested in measuring the trend in $\Gamma_{\rm ML}$ with $T_{\rm hot}$, as opposed to its absolute value.  

\begin{deluxetable*}{LLLLLLLLLLhLL}
\def\arraystretch{1}
\tabletypesize{\footnotesize}
\tablecaption{\ovi emission in mixing layer models compared to cooling flow models at various hot phase temperatures.}
\label{tab:ml}
\tablehead{
    \colhead{$T_{\rm hot}$} & \colhead{$P_{\rm hot}/k_B$} & \colhead{$Z$} & \colhead{$h$} & \colhead{$\mathcal{M}_{\rm rel}$} & \colhead{$f_\nu$} & \colhead{$\kappa_0$} & \colhead{${\rm Pr}$} & \colhead{$\tau$} & \colhead{$\dot{m}$} & \nocolhead{$F_{\rm O\,VI}$} & \colhead{$\Gamma_{\rm ML}$} & \colhead{$\Gamma_{\rm ML}/\Gamma_{\rm CF}$} \\
    \colhead{(K)} & \colhead{(K cm$^{-3}$)} & \colhead{($Z_\odot$)} & \colhead{} & \colhead{} & \colhead{} & \colhead{} & \colhead{} & \colhead{} & \colhead{($M_\odot\,{\rm yr}^{-1}\,{\rm kpc}^{-2}$)} & \nocolhead{(erg s$^{-1}$ cm$^{-2}$)} & \colhead{(erg g$^{-1}$)} & \colhead{}
}
\startdata
\hline
5 \times 10^6 & 1000 & 0.3 & 1.0 & 1.0 & 0.053 & 10^{-6} & 5.4 \times 10^{-4} & 0.1 & 4.5 \times 10^{-5} & 3.0 \times 10^{-9} & 1.0 \times 10^{13} & 0.61 \\
1 \times 10^7 & 1000 & 0.3 & 0.7 & 0.7 & 0.053 & 10^{-6} & 4.1 \times 10^{-4} & 0.1 & 2.8 \times 10^{-5} & 2.9 \times 10^{-9} & 1.5 \times 10^{13} & 0.92 \\
3 \times 10^7 & 1000 & 0.3 & 0.4 & 0.4 & 0.053 & 10^{-6} & 2.5 \times 10^{-4} & 0.1 & 1.3 \times 10^{-5} & 2.6 \times 10^{-9} & 3.0 \times 10^{13} & 1.82 \\ 
6 \times 10^7 & 1000 & 0.3 & 0.3 & 0.3 & 0.053 & 10^{-6} & 1.8 \times 10^{-4} & 0.1 & 0.8 \times 10^{-5} & 2.5 \times 10^{-9} & 4.8 \times 10^{13} & 2.86 \\
\hline
\enddata
\end{deluxetable*}


The results of this simulation set are presented in Table \ref{tab:ml}.  We immediately notice that $\Gamma_{\rm ML}$, and thus $\Gamma_{\rm ML}/\Gamma_{\rm CF}$, rise monotonically by a factor of $\sim 5$ with temperature within the range probed.  This trend is largely driven by a decrease in the mass flux through the layer, $\dot{m}$, while the \ovi emission itself ($\dot{m}\Gamma_{\rm ML}$) largely remains relatively constant.  The mass flux must decrease to maintain the boundary conditions (${\rm d}T/{\rm d}z \rightarrow 0$ at both the hot and cold ends) and keep the mixing layer stable.  We note that $\Gamma_{\rm CF}$ itself also stays roughly constant throughout the temperature range probed, as mentioned above.  Our choice to maintain constant physical conditions, as opposed to self-similarity, is the most conservative choice for this comparison, as maintaining a constant $h$ and $\mathcal{M}_{\rm rel}$ would impose a physically wider mixing layer with faster shear velocities at higher $T_{\rm hot}$, increasing the emitting volume for \ovi and further exacerbating the observed trend in $\Gamma_{\rm ML}$.

Given that $\Gamma_{\rm ML}(T_{\rm hot})$ monotonically increases, it is apparent that changing the hot-phase temperature alone in these steady 1D mixing layer models is not sufficient to reproduce the observed trend of increasing $\sfr/\mdotovi$ with $\mdotovi$---the hotter temperature models, applicable to the high-mass end, would further pull systems to the left, whereas the cooler temperature models, applicable to the low-mass end, would pull systems to the right, with the overall effect of steepening the observed trend and making the tension worse.  However, we refrain from claiming that this rules out mixing layer models entirely.  In addition to all of the caveats on these models already mentioned, another potentially important factor is the physical 1D structure of the layer itself, which may be too simplistic to provide an adequate representation of the complex 3D structure of real cluster atmospheres, with cold clouds embedded in a volume-filling hot atmosphere, and a time-dependent multiphase feedback cycle.  In more recent work, \citet{2026arXiv260607741C} have acknowledged that this simplification causes the X-ray/H$\alpha$ ratio to be discrepant from observations by up to an order of magnitude, and found that properly accounting for the clumpy structure of the gas recovers the correct ratio.  Therefore, it is reasonable to expect that a more careful treatment of \ovi emission, within feedback-regulated hydrodynamic simulations, and integrated over synthetic COS-sized apertures, might change the distribution of \ovi emission per unit mass over halo temperature substantially.


\section{Conclusion} \label{sec:conclusion}

We have presented a combination of new and archival observations of massive starbursting BCGs with \textit{HST}'s Cosmic Origins Spectrograph, measuring star formation rates and \ovi-based cooling rates, along with maximal cooling rates from archival \textit{Chandra} data, and X-ray spectroscopic cooling rates gathered from the literature.  We have used these data to extend the dynamic range of the ``Reigniting the \textit{FUSE}'' sample, which consists of 29 massive elliptical galaxies with archival data from the Far-Ultraviolet Spectroscopic Explorer (Paper I and Paper II).  

The star formation rates we measure are in good agreement with previous values from the literature, and our parametric star formation history model strongly suggests that this is indicative of active ongoing star formation, as opposed to post-starburst signatures.

The new data confirm the previously established trends in \ovi-based cooling efficiency continue across an additional decade in mass, being suppressed by a factors of 4--5 ($\mdotovi/\mdotmaxap$) and 8--10 ($\mdotovi/\mdotmax$) in massive clusters relative to lower mass groups. The difference is attributed to qualitatively distinct forms of AGN and SNe feedback, acting on different radial scales and driving inflows and outflows of gas at different rates.

In this subsample of some of the strongest cool cores known, traced by the new COS measurements (and also including Perseus from \textit{FUSE}), we find that the X-ray spectroscopic cooling rates \mdotx, \ovi spectroscopic cooling rates \mdotovi, and star formation rates all approach each other, while retaining an object-to-object scatter exceeding that expected purely from statistical measurement uncertainty. This puts these systems in a separate category from the remainder of the \textit{FUSE} sample, which overwhelmingly show $\sfr/\mdotovi \sim \sfr/\mdotx \sim 10\%$.  As a result, we infer that the enhancement in $\sfr/\mdotmax$ that is seen in strong cool cores is a result of an increase in the conversion efficiency of warm \ovi-emitting gas into stars ($\sfr/\mdotovi$), as opposed to an increase in the cooling efficiency of the hot ICM into the \ovi phase ($\mdotovi/\mdotmax$).  This could be explained by (1) a transition in the dominant physics between mixing layers and cooling from low- to high-mass halos; or (2) time lags and hysteresis between \ovi emission and star formation, driven by molecular gas accumulation and cloud recycling, which smoothly converge at larger \sfr and \mdotovi due to faster depletion and replenishment times for the molecular gas.  However, these interpretations remain subject to uncertainties, such as potential sample selection biases between the \textit{FUSE} and COS samples, and the effect of the varying molecular H$_2$ masses on the depletion/replenishment timescale argument. Spatially resolved \ovi kinematics, star formation histories, and molecular gas masses could help distinguish which of these pictures is most accurate.



These results demonstrate the value of adding new data and increasing a sample's dynamic range, while also highlighting the importance of considering sample biases when studying any sample with a complicated selection function.  A clear necessity is demonstrated for more far-ultraviolet spectroscopy of massive elliptical galaxies at lower star formation rates, to ascertain the unbiased trends in \sfr, \mdotovi, \mdotmax, and \mdotx.  Current (\textit{HST}-COS) and next-generation (\textit{Aspera}, \textit{UVEX}, \textit{HWO}) telescopes with far-ultraviolet capabilities are absolutely critical for this science moving forward.

\begin{acknowledgements}

The authors humbly thank Greg Bryan for thoughtful discussions about this work's results and methodology and statistical analysis.


MR acknowledges support from the National Science Foundation Graduate Research Fellowship under Grant No. 2141064.  


MR and MM received funding for this project from the NASA Astrophysics Data Analysis Program under grant 80NSSC25K7560.


MC acknowledges support from NASA grant 80NSSC23K0461.

MG acknowledges support from the ERC Consolidator Grant \textit{BlackHoleWeather} (101086804).

VO acknowledges support from DICYT ESO-Chile Comite Mixto PS 1757, FONDECYT Regular 1251702, CIRAS-AI Project, code FIUF137139-USACH.


This research is based on observations with the NASA/ESA Hubble Space Telescope and Far Ultraviolet Spectroscopic Explorer, obtained from the MAST data archive at the Space Telescope Science Institute, which is operated by the Association of Universities for Research in Astronomy, Inc., under NASA contract NAS5-26555. Support for Program number HST-GO-17716 was provided through a grant from the STScI under NASA contract NAS5-26555.  All of the {\it HST} data used in this paper can be found in MAST: \dataset[10.17909/2yrf-1m26]{http://dx.doi.org/10.17909/2yrf-1m26}.  All of the {\it FUSE} and {\it GALEX} data used in ``Reigniting the FUSE'' can also be found in MAST: \dataset[10.17909/mgq9-2395]{http://dx.doi.org/10.17909/mgq9-2395}.


The scientific results reported in this article are based in part on data obtained from the Chandra Data Archive.  


This research has made use of the NASA/IPAC Extragalactic Database: \dataset[10.26132/NED1]{http://dx.doi.org/10.26132/NED1}, which is funded by the National Aeronautics and Space Administration and operated by the California Institute of Technology.

\end{acknowledgements}

\appendix 

\begin{figure*}
    \centering
    \includegraphics[width=0.39\linewidth]{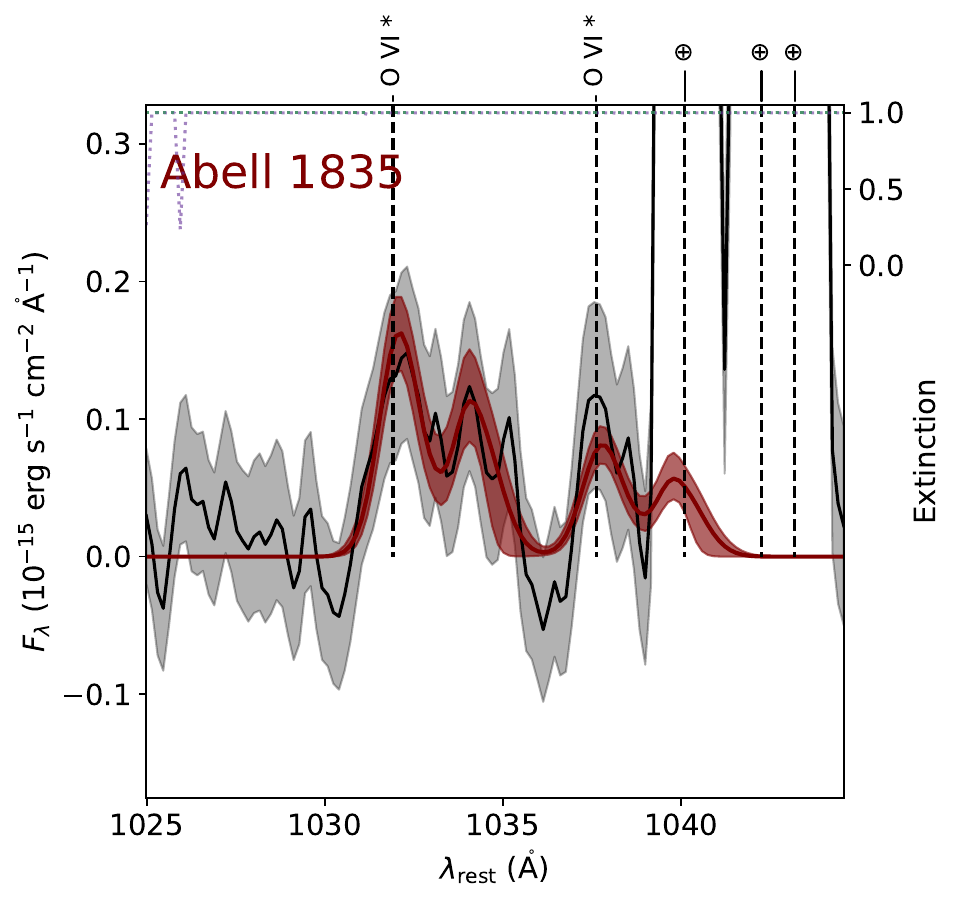}
    \includegraphics[width=0.39\linewidth]{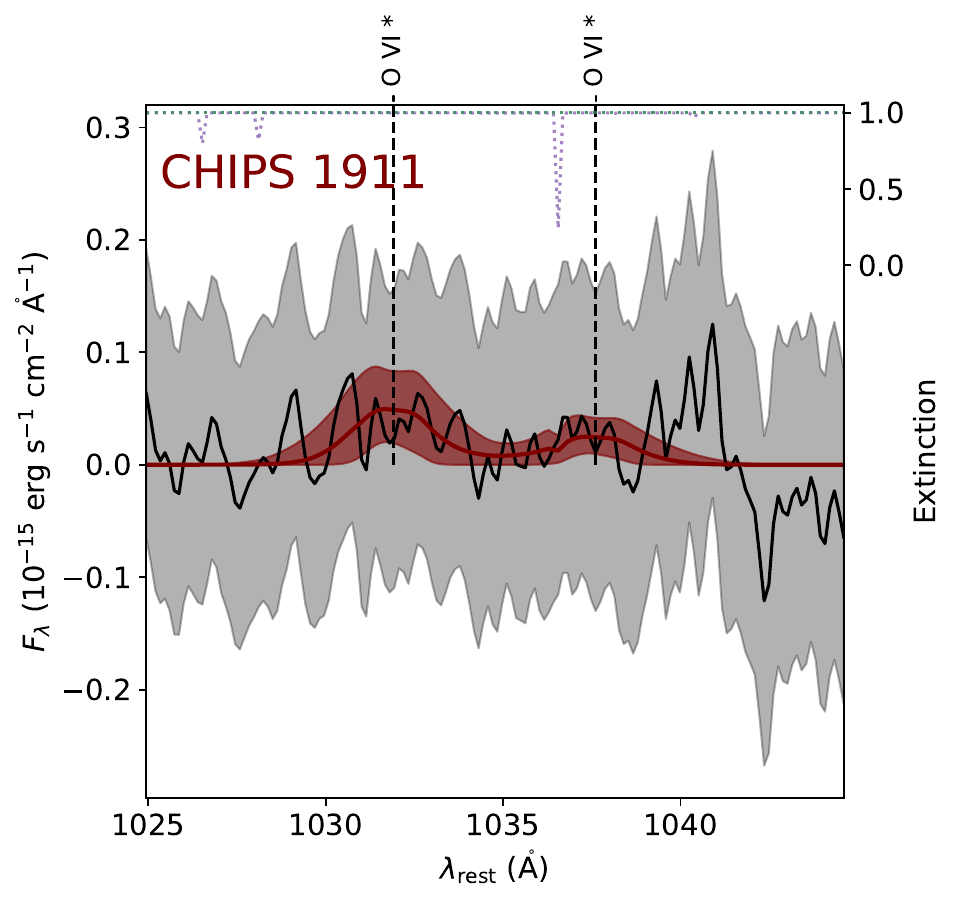}
    \includegraphics[width=0.39\linewidth]{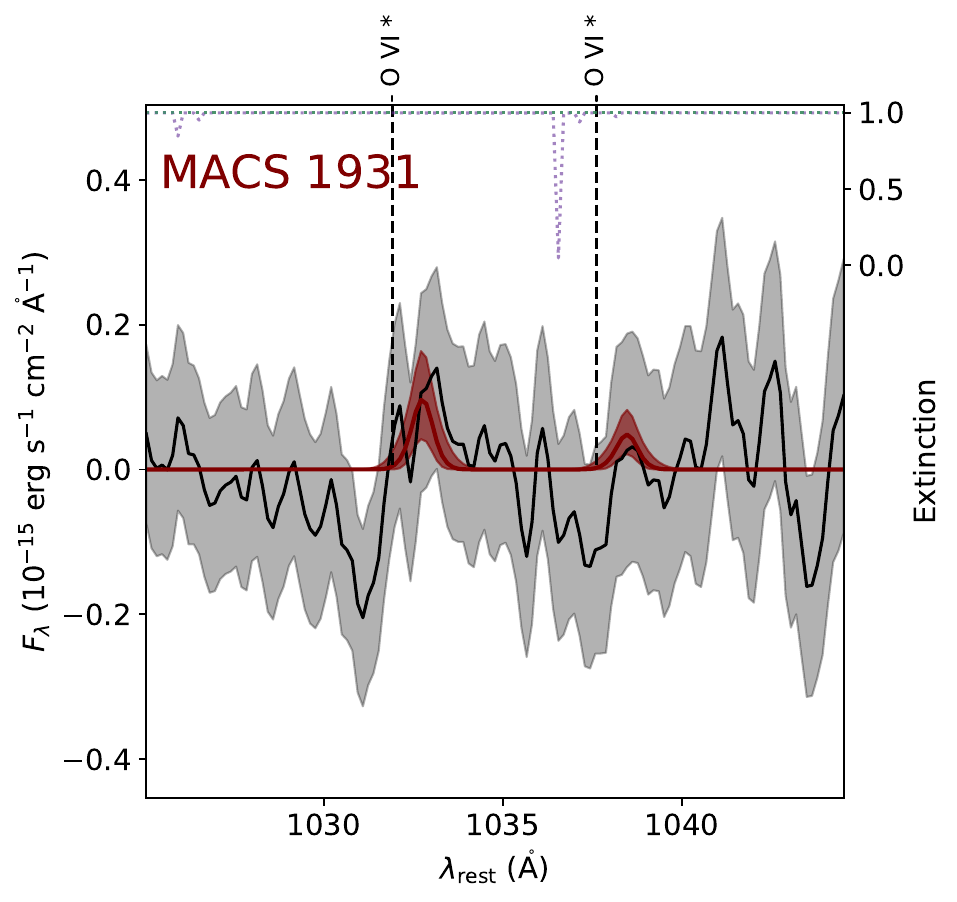}
    \includegraphics[width=0.39\linewidth]{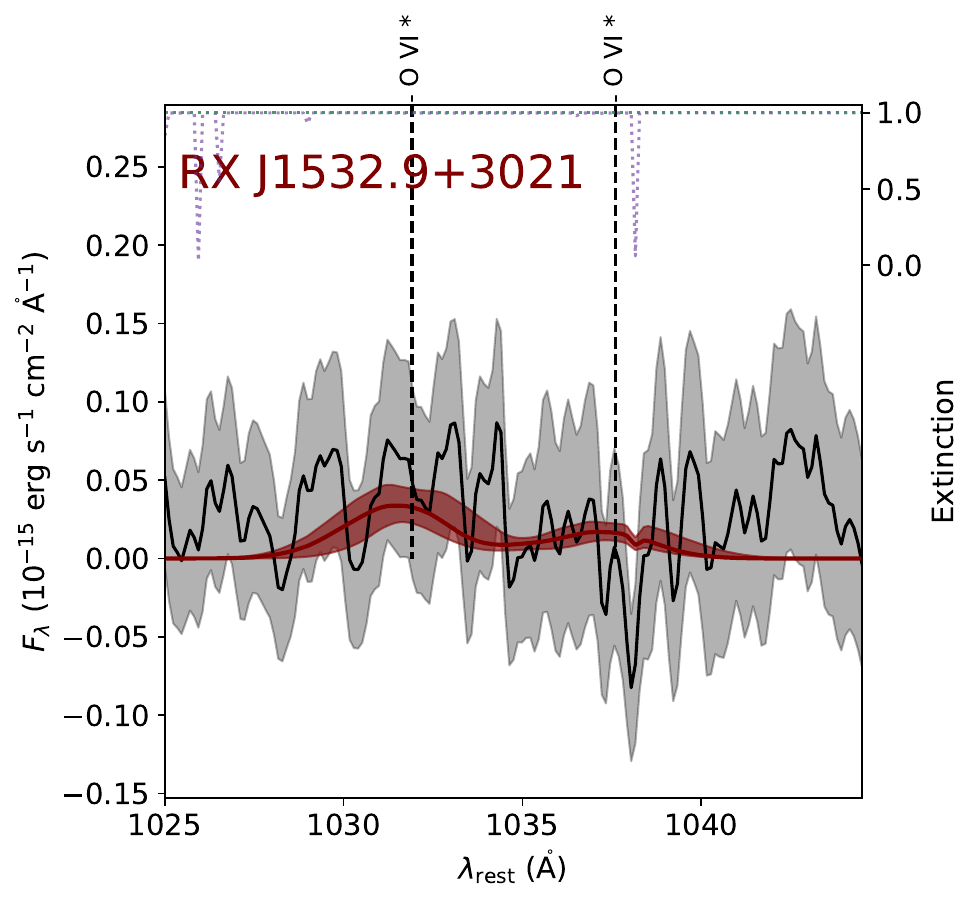}
    \includegraphics[width=0.39\linewidth]{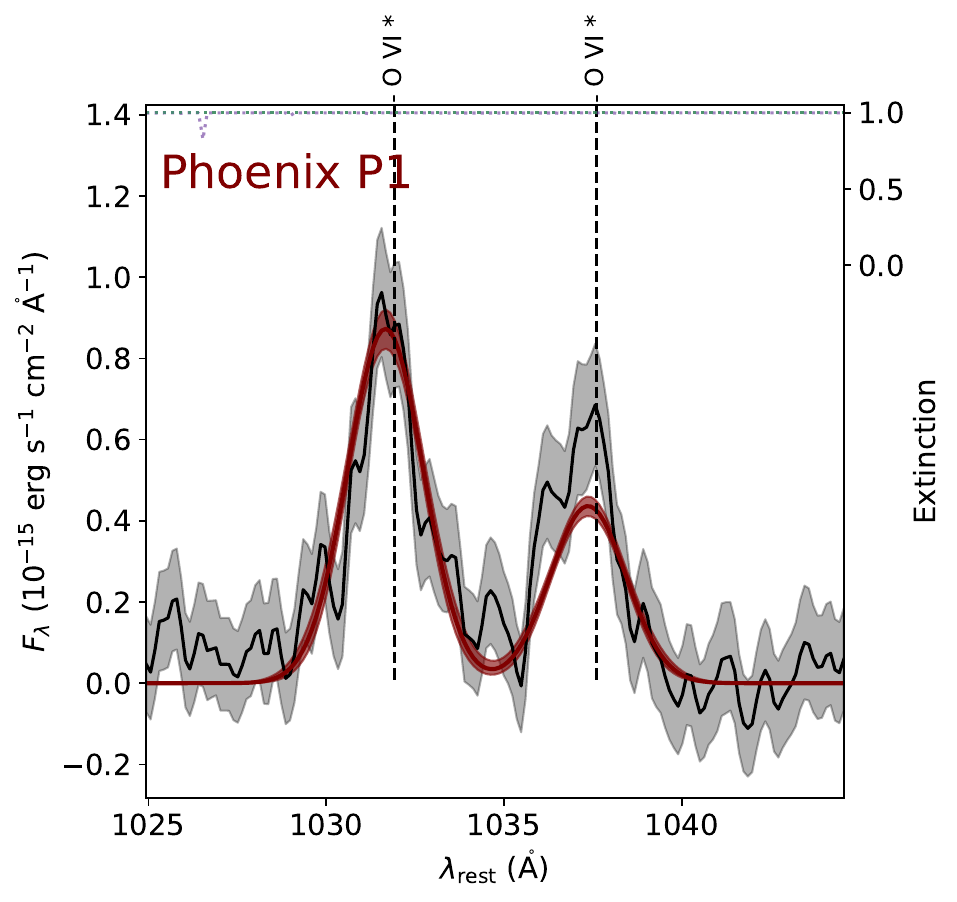}
    \includegraphics[width=0.39\linewidth]{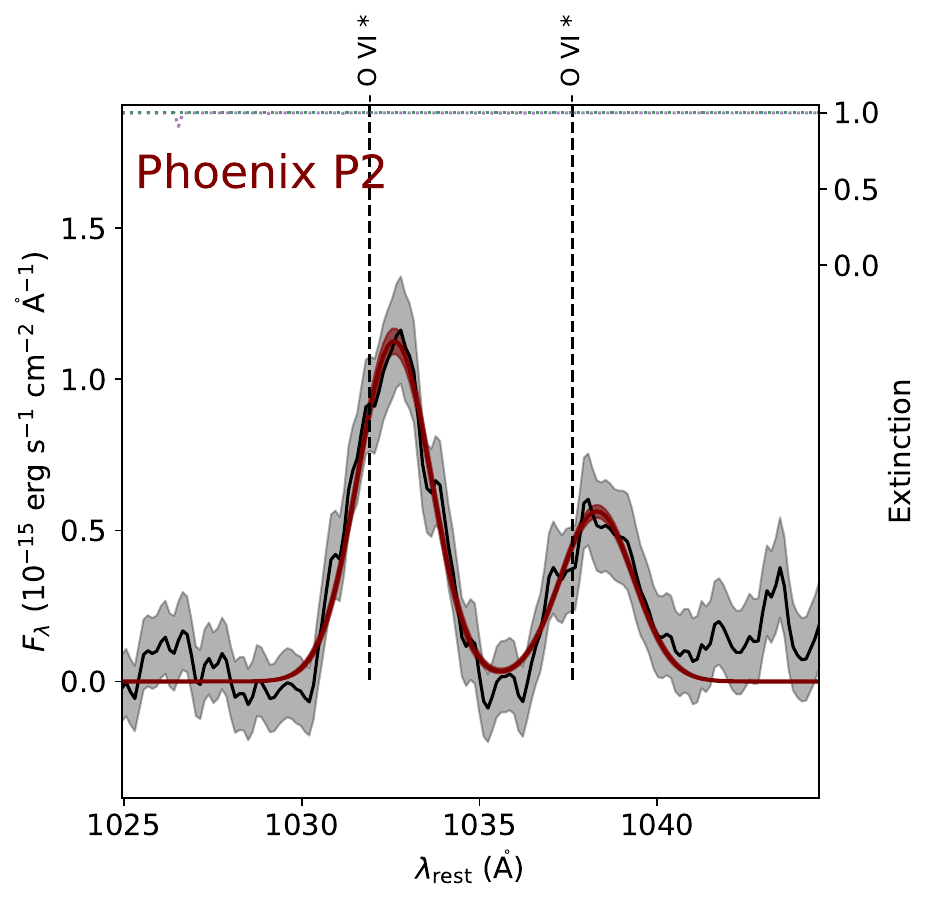}
    \caption{Plots show fits to the \oviwave doublet for each system.  The continuum-subtracted spectra are shown in black, with a corresponding gray shaded area showing the $\pm 1\sigma$ uncertainty region.  The model is shown in orange, with a corresponding orange shaded area showing the $\pm 1\sigma$ uncertainty region on the model from bootstrapping.  The \ovi lines are labeled with vertical dashed lines, while airglow lines are labeled with $\oplus$.  Abell 1835 uniquely fits the \ovi lines with two velocity components, as explained in the text in $\S$\ref{sec:results}.   For visual clarity, the data and models have both been smoothed by a Gaussian kernel with $\sigma = 1$ pixel ($0.2~\angstrom$), while the unsmoothed data are used during the fitting process.}
    \label{fig:ovi_fits}
\end{figure*}

Plots of each system's spectral fits to the \oviwave doublet are given in Figure \ref{fig:ovi_fits}.

\bibliography{main}{}
\bibliographystyle{aasjournalv7}

\end{document}